\documentclass[]{aa}
\usepackage{natbib}
\usepackage{epsfig}
\usepackage{graphicx}
\usepackage{txfonts}
\usepackage{times}

\newcommand{\sub}[1]{_{\rm #1}}
\newcommand{\reference}[1]{}
\newcommand{\changed}{}
\newcommand{\newchanged}{}
\def\apj{{ApJ }}
\def\ApJ{{ApJ }}

\def\AandA{{A\&A }}

\def\aap{{A\&A }}

\def\mnras{{MNRAS }}

\begin{document}

\title{Interstellar cloud structure: The statistics of centroid
velocities}
\author{V.~Ossenkopf\inst{1,2}, A. Esquivel\inst{3}, A. Lazarian\inst{3}, \and
J. Stutzki\inst{1}}
\institute{I. Physikalisches Institut der Universit\"at zu K\"oln,
Z\"ulpicher Stra\ss{}e 77, 50937 K\"oln, Germany
\and 
SRON National Institute for Space Research,  P.O. Box 800, 
9700 AV Groningen, the Netherlands
\and
Astronomy Department, University of Wisconsin-Madison, 475 N.
Charter St., Madison, WI 53706, USA}

\titlerunning{The statistics of centroid
velocities}
\authorrunning{Ossenkopf et al.}

\date{Received: January 24, 2005 / Accepted: January 25, 2006}

\abstract
{The investigation of the statistical properties of maps
of line centroids has been used for almost 50 years, but there is still
no general agreement on their interpretation.\\
{\bf Aims:}
We try to quantify which properties of underlying turbulent velocity
fields can be derived from centroid velocity maps, and we test conditions
under which the scaling behaviour of the centroid velocities matches the
scaling of the three-dimensional velocity field.\\
{\bf Methods:}
Using fractal cloud models we study systematically 
the relation between three-dimensional density and velocity fields
and the statistical properties of the produced line centroid maps.
We put special attention to cases with large 
density fluctuations resembling supersonic interstellar turbulence.
Starting from the $\Delta$-variance analysis we derive a new 
tool to compute the scaling behaviour of the three-dimensional velocity
field from observed intensity and centroid velocity maps.\\
{\bf Results:}
We provide two criteria to decide whether the information from the centroid velocities 
directly reflects the properties of the underlying velocity field.
Applying these criteria allows to understand the different results found 
so far in the literature on the interpretation of the statistics of
velocity centroids. The new iteration scheme can be used to derive 
the three-dimensional velocity scaling from centroid velocity maps
for arbitrary density and velocity fields, but it requires an accurate 
knowledge of the average density of the considered interstellar cloud.
}

\maketitle

\section{Introduction}

Understanding the role and nature of interstellar turbulence
has been the subject of intensive studies for half a century now
but still remains open in many aspects \citep[cf.][]{Elmegreen}.
Major questions concern the mechanisms by which turbulent
motions are driven and the role of the strong compressibility
of the interstellar medium on the structure of the turbulent
energy cascade. Both aspects are directly reflected on the
spectrum of velocity fluctuations in the turbulent motion.
It is frequently claimed, that
driving mechanisms should create dominant motions at the
corresponding scales, and the power spectrum of velocities
in the turbulent cascade is known to change from a
$P(|\vec{k}|)\propto |\vec{k}|^{-11/3}$
Kolmogorov spectrum for an incompressible medium to a
$P(|\vec{k}|)\propto |\vec{k}|^{-4}$ spectrum of Burger's
turbulence in a highly compressible medium dominated by shocks
\citep{Chappell}. {\changed However, numerical simulations show
often a different behaviour \citep[see][]{Cho2005}
which makes the issue of the observed spectrum very intriguing.}

To support the theoretical understanding of the interstellar
turbulence it is thus essential to actually measure the velocity
structure in the interstellar medium. Unfortunately, there
is no direct way to do so.
Observations of the profiles of atomic or molecular lines from
interstellar clouds allow to deduce information on the line-of-sight
velocity structure of the clouds. 
{\changed The problem of recovering of the velocity information from lines
is far from being straightforward.  Even} in the most simple
case of thermally excited optically thin lines from an isothermal
medium the line profiles originate from a convolution of the density
structure $\rho$ depending on the sky coordinates $\vec{x}=(\alpha,\delta)$
and the line-of-sight coordinate $z$ with the velocity structure
$v_z(\vec{x},z)$:
\begin{equation}
I(\vec{x},v) \propto \int dz\; \rho(\vec{x},z)
\phi(v-v_z(\vec{x},z)) 
\label{eq_zconvol}
\end{equation}
In the limit of narrow lines, the line profile $\phi(v-v_z(\vec{x},z))$
can be approximated by a $\delta$-function. 
{\changed There are several complementary ways to use this information 
\citep[cf.][]{Lazarian2004}. Here, we restrict ourselves to centroids, the first
moment of the lines,} but the centroids still provide no direct map of
the velocity structure.

Models for the density structure and the relation between density
and velocity structure are needed to deduce the latter from the profiles
$I(\vec{x},v)$. This is straight-forward for simple geometries
like spherical clouds or thin disks but extremely difficult for 
filamentary turbulent cloud structures showing varying 
substructures on all spatial scales. Hydrodynamic
or magneto-hydrodynamic numerical simulations can be used as physically
justified models for turbulent interstellar clouds within a limited
dynamic range.  The nature of these can be described, however,
only in terms of statistical measures. Fractal cloud models provide
a reasonable phenomenological description of the clouds. We focus on measures
for the spatial scaling of the velocity structure. The ultimate
goal is to derive the three-dimensional (3-D) power spectrum
of velocity fluctuations.

{\changed A recovery of 3-D information from the available 2-D data
requires, in general, an inversion, which may result in substantial noise
in the inverted data. For deriving the turbulence statistics we can,
however, use its symmetries. Here, we restrict ourselves to statistically
isotropic turbulence. The derivation of properties of anisotropic, but
axisymmetric turbulence from observations was discussed by 
\citet{Lazarian1995}. Anisotropies can be due to the magnetic
fields \citep{Higdon,Zank,Goldreich}. However, if, as both
theory and numerics suggests \citep[see][]{Goldreich,Cho2003}, the
energy spectrum is dominated by fluctuations perpendicular
to the local direction of magnetic field, the effects of anisotropy
on the observed spectra can be neglected \citep{Esquivel03}.

In order to derive the isotropic power spectrum} we will use
an auxiliary quantity, the $\Delta$-variance spectrum, 
because of practical advantages
when measuring the velocity scaling in observed data. Moreover,
we restrict the analysis here to the first moments of the lines, the
centroid velocity, as the most obvious tracer to measure the
velocity structure in an interstellar cloud.

Maps of observed line centroids have been systematically studied
to obtain the scaling behaviour of centroid velocity
differences as a function of lag for almost 50 years now
\citep[e.g.][]{Muench, Kleiner, Miesch94, Lis96, 
Miesch99}. However, there is still no agreement on the
theoretical relation between the observed scaling behaviour of the
centroid velocities and the scaling behaviour of the underlying
turbulent velocity structure. {\changed
Although it was clear from the very beginning that density structure
can influence the line centroids, up to {\newchanged the recent past}
there was no criterion to estimate to quantitative effect of density.}

Investigating hydrodynamic turbulence simulations \citet{OML} found
that the centroid maps show approximately the same Hurst index, i.e.
the same relative variation across a given scale,  as the underlying
3-D velocity structure. This means that the power spectral index in their
centroid maps was reduced by one compared to the power spectral index in
the 3-D velocity structure. Studies of fractal clouds
by \citet{Miville} showed in contrast that their centroid maps show
the same power spectral index as the 3-D 
velocity structure\footnote{{\changed When dealing with projected quantities
one has to carefully distinguish correlation functions and power spectra.
When a power-law approximation is good for both of them, the spectral index
of correlation functions gets steeper by one due to projection, while the 2-D
projected power spectrum retains the spectral index of the underlying
3-D spectrum.}} {\changed \citet{Lazarian03} provided an analytical
treatment of the centroid statistics introducing a
new more robust definition of velocity centroids, formulated
a criterion when the centroids represents the velocity
statistics, but this publication did not cover
the  parameter space to be fairly compared with previous studies.
The problem was further elaborated in a subsequent study by
\citet{Levrier}, who pointed out that the statistical treatment 
presented in the form of structure functions by \citet{Lazarian03} may 
have some advantages if rewritten in terms of correlation functions. 
Assuming that the fluctuations are small compared to the mean density
he obtained analytic expressions for correlation functions of centroids.
Combining structure and correlation functions \citet{Esquivel} provided
a detailed study of centroid velocities for data obtained through 
compressible MHD simulations.}
Here, we compare the different centroid definitions and test their outcome
for a set of fractal cloud models. 

Using the $\Delta$-variance analysis of the centroid maps, we show
that it is in principle applicable to derive the velocity power spectrum from
observed centroid maps but that the reliability of this derivation depends
critically on individual turbulence parameters. The centroid maps
reflect the actual velocity distribution only in a medium with an
average density which is large compared to the density dispersion.
Here, the $\Delta$-variance analysis
provides a direct measure for the power spectral index of the
velocity structure. Only when applied in an iterative process with
an a priori knowledge of the average density, the analysis of centroid
maps allows to approximate the velocity structure in the general case.
The approximation is better as steeper the velocity spectrum is and
as better the average density is known.

In Sect.~2 we shortly repeat the formalism used to describe
the velocity centroids, discuss the 
properties of the test data sets, 
and the ways to measure their spatial scaling behaviour in terms
of the $\Delta$-variance. 
In Sect.~3 we perform the analysis of the centroid maps using
the $\Delta$-variance, compare the results with the original test data
and derive criteria when the centroid maps can be used to measure
directly the three-dimensional velocity structure.
In Sect.~4 we propose an iterative method to derive the 
power spectrum of the velocity structure from the centroid maps
in cases without a direct matching.
Sect.~5 gives a summary
with respect to the interpretation of observed data.

\section{The starting point}

\subsection{Definition of centroid velocities}

For the fluctuating density and velocity fields in a cloud we can always
write
\begin{eqnarray}
\rho(\vec{x},z)&=&\rho_0+\delta\rho \nonumber \\
v(\vec{x},z)&=&v_0+\delta v 
\end{eqnarray}
where $\rho_0$ and $v_0$ are averages over the whole cloud, and $\delta\rho$
and $\delta v$ denote the variations across the cloud\footnote{From here on
we drop the index $z$ in the notation for the line-of-sight component of the velocity
because we consider only this component.}.

When we assume that the emissivity is proportional to the density of the
cloud, the line intensity $I(\vec{x},v)$ at velocity $v$ is a
measure for the total column density of emitters with this velocity at a given
line-of-sight $\vec{x}$. This condition is violated for optically
thick lines or media with strongly varying temperatures but it is e.g.
well fulfilled for the [C{\sc II}] emission from the cold
neutral medium or the H{\sc I} emission from the warm neutral medium.
The effect of self-absorption will be quantified in a subsequent paper.
For constant emissivity the integrated line intensity is 
\begin{eqnarray}
I\sub{int}(\vec{x}) &=& \int dv\; I(\vec{x},v) \nonumber\\
&=& X \int dz\; \rho(\vec{x},z) 
\end{eqnarray}
where $X$ is the proportionality factor from Eq.~(\ref{eq_zconvol})
translating the column density into a line intensity.

There are two different centroid definitions in common use.
Ordinary centroid velocities, also known as normalised centroids,
are obtained as
\begin{eqnarray}
v\sub{c,norm}(\vec{x}) &\!\!=\!\!& \int dv\; v I(\vec{x},v) \bigg/ \int dv\; I(\vec{x},v) 
\\
&\!\!=\!\!&\int dz\; (v_0+\delta v)(\rho_0+\delta \rho) \bigg/ \!\int dz\; (\rho_o +\delta \rho)
\nonumber
\end{eqnarray}

Unfortunately, this definition implies a complex combination of
density and velocity fluctuations which makes it impossible to 
disentangle the influence from both structures in the general
case. Only in case of very small fluctuations, a linearisation
technique can be developed \citep{Levrier}.
A better separation  of density and velocity fluctuations in the
centroids is obtained when we apply the definition of weighted, i.e.
unnormalised, centroids as proposed by \citet{Lazarian03}\footnote{In
contrast to the original definition we have not included the
constant factor $X$ in the centroid definition so that
the weighted centroids have the dimension of a velocity times
column density here instead of velocity times intensity. This keeps the
equations in the following sections somewhat shorter.}
\begin{eqnarray}
v\sub{c}(\vec{x}) &=& 1/X \int dv\; v I(\vec{x},v) 
\nonumber \\
&=& v_0 \rho_0 z\sub{tot} + \rho_0 \int dz\; \delta v \nonumber\\
&& + v_0 \int dz\; \delta \rho + \int dz\; \delta \rho \delta v 
\label{eq_centroids}
\end{eqnarray}
where $z\sub{tot}$ is the total thickness of the cloud. In
this definition the centroids do not have the dimension of a
velocity but of velocity times column density. For a better
comparison with the ordinary centroid velocities it is useful to
normalise the weighted centroids by the average column density
$\rho_0 z\sub{tot}$, but we omit this factor in the following
to keep the equations shorter. The constant factor would
not change any of our conclusions on the scaling behaviour
of the velocity structure.

We see that even in this definition the centroid velocities
are not simply determined by the projected velocities $v_0+\int dz\; \delta v$
but also by two terms reflecting the density variations. The contribution 
from the projected density variations $\int dz\; \delta \rho$ can be easily 
obtained from the integrated line profiles and it can be 
eliminated by selecting a velocity scale with $v_0=0$. However,
the term containing the product of the fluctuations in the density
and the velocity structure cannot be measured separately. 

The scaling behaviour of the centroid velocities
depends on the combination of density and velocity
variations along the line of sight, which cannot be retrieved directly.
The relative contribution of the simple projection of the
velocity structure and the density variations $\delta\rho$ across
the line of sight 
depends on the ratio between the density fluctuations $\delta \rho$
and the average density $\rho_0$. Eq. (\ref{eq_centroids}) thus
shows already that the ratio between the density dispersion
$\sigma_{\rho}$ and the average density $\rho_0$ is a critical
parameter for the relation between the 3-D velocity scaling and
the centroid scaling.

\subsection{Test data sets}
\label{sect_fbms}

To study the general ability of different methods to extract
the underlying velocity structure from observed centroid velocities
we construct well defined test data sets for the density
and velocity structure which are used
to study the translation of their scaling properties
into centroid properties.

Interstellar cloud observations often reveal self-similar scaling
properties \citep[e.g.][]{Falgarone, Combes} corresponding to power-law
power spectra of the intensity distribution. Such intensity
maps can be approximately modelled by 
fractional Brownian motion (fBm) structures \citep[see e.g.][]{Stutzki98,
Bensch}. They are defined by  the single number $\beta$ determining
the exponent of the power spectrum, $P(|\vec{k}|)\propto |\vec{k}|^{-\beta}$.
The phases of the Fourier spectrum are random.

Thus fBm's represent one of the simplest possible representations of
interstellar cloud structures still allowing a parameter study
in terms of the spectral index $\beta$ which determines the
actual appearance of the structures. fBm's can be defined in
arbitrary dimensions and we use their essential property
that the projection of an fBm to lower dimensions results in
a new fBm with the same spectral index \citep{Stutzki98,Brunt}.\footnote{As
discussed by \citet{Stutzki98} {\changed it is easy to show that this fBm
property violates the often used hypothesis} that the fractal dimension
decreases by one in projection \citep{Peitgen}.} {\newchanged Thus
the spectral index measured for the column density directly reflects 
the index of the three-dimensional density structure.}

Measured spectral indices for the {\newchanged column} density structure of
interstellar
clouds range from {\changed 2.0 to 3.7 \citep{Elmegreen,Falgarone04}. 
Observations of large molecular clouds and molecular cloud complexes 
and H{\sc I} absorption line studies provided typical values 
between 2.4 and 2.9  \citep[e.g.][]{Stenholm, Langer, Deshpande, Bensch,
Huber, Padoan2003}, whereas
\citet{Bensch} found indications for somewhat larger indices at
the scales of cloud cores. Observations of the warm atomic gas provided
typical values between 3.3 \citep[e.g.][]{Snezana} and
3.6 \citep{Miville2} with some indications for an even broader range
from 8/3 to 11/3 in the LMC {Elmegreen2001}.
Due to a lack of direct measurements, as discussed in the introduction,
the index range of the velocity structure is still hardly known.}
MHD simulations by \citet{Cho2003} indicate that it should be close to
the Kolmogorov value of 11/3.
{\changed In contrast \citet{OML} and \citet{BruntHeyer} obtained
velocity spectral indices close to four from observations of 
the Polaris Flare molecular cloud and of molecular clouds in the FCRAO survey
of the Outer Galaxy, respectively, consistent with the properties of a
shock-dominated medium. In these cases, the velocity spectrum was
always steeper than {\newchanged column} density spectrum.
Here, we do not aim at reproducing the exact combination of spectral
indices for any particular interstellar cloud, but want to study the
general behaviour covering the full range of spectral indices observed so far.}

\citet[][]{Esquivel} demonstrated that the centroid
structure function shows a qualitatively different behaviour for spectra
with an index above and below 3.0 (steep and shallow spectra).
{\changed Unfortunately, the observational data do not rule out
either of the two types. Thus we focus on two test data sets:
fBm's with a spectral index of 3.7 representing steep spectra and 
with an index of 2.6 representing a shallow behaviour}, respectively.
They sample both regimes and are close to {\changed some observed values
for the velocity and density structure.}
We have studied a much larger parameter range covering spectral
indices between 2.0 and 4.0 but with the four
possible mutual combinations of the two mentioned spectral indices all
major effects are covered so that we restrict ourselves
to these cases for all examples given in the following.

\begin{figure}
\centering
\epsfig{file=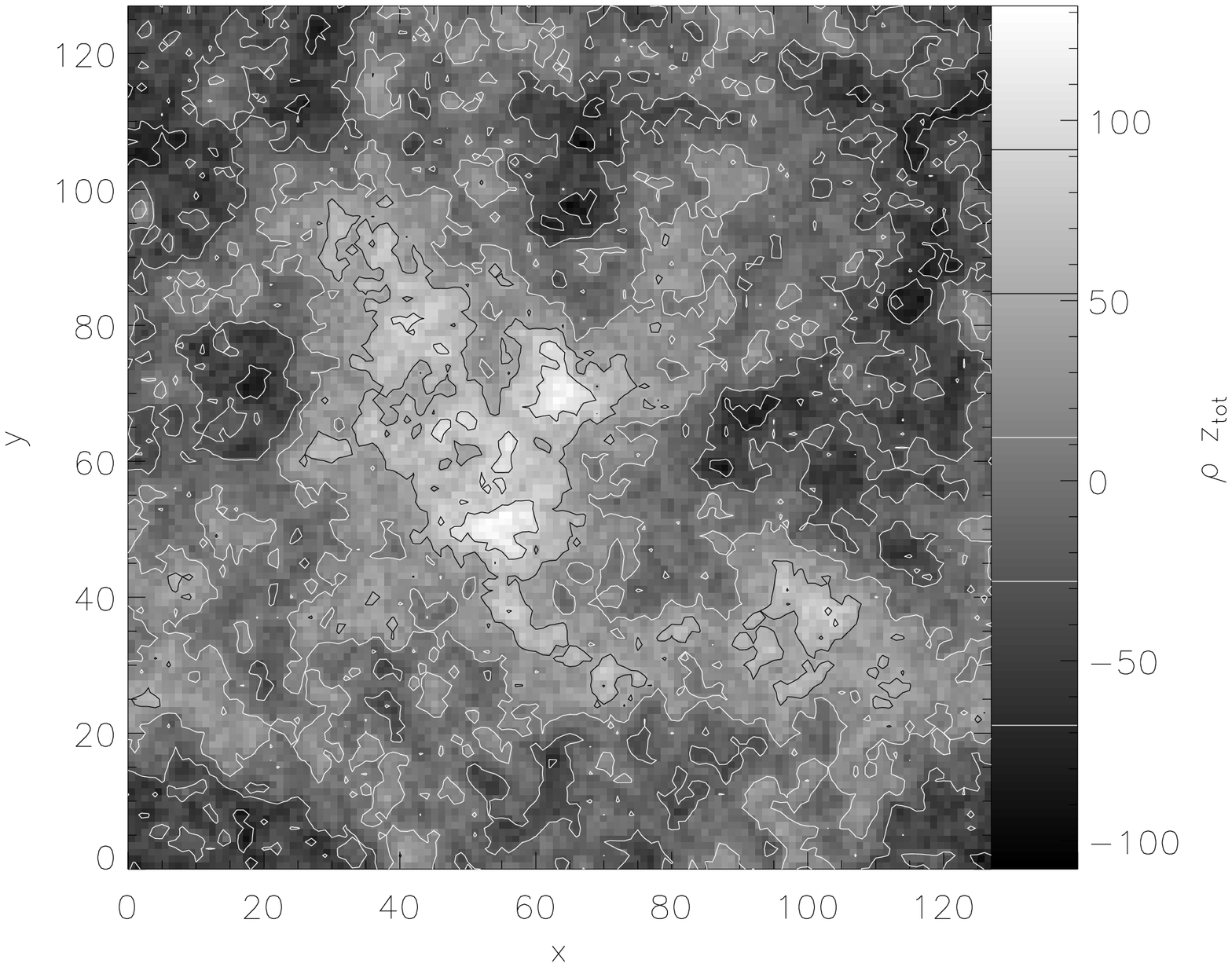,angle=90,width=\columnwidth}\\
\epsfig{file=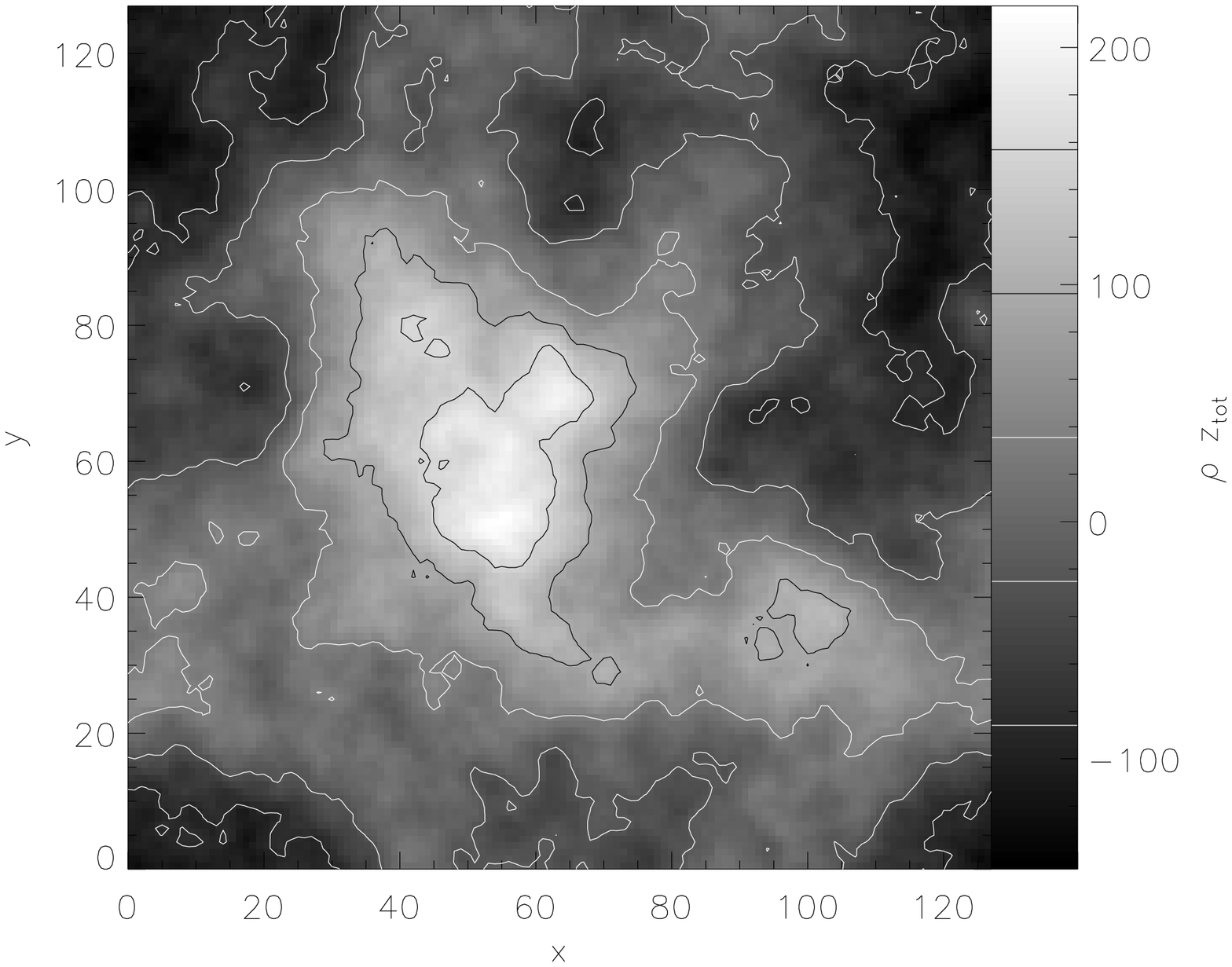,angle=90,width=\columnwidth}
\caption{Projected maps of fBm structures with 
spectral indices $\beta=2.6$ (upper plot) and $\beta=3.7$
(lower plot). Both data sets use the same random phases leading
to the apparent similarity of the overall distribution
in this example.
}
\label{fig_densitymaps}
\end{figure}

In Fig. \ref{fig_densitymaps} we give a visual impression for
the difference in the actual projected structure between fBm's 
of different spectral index. The spectral index basically
determines the relative contribution of structures on different
size scales. The fBm with an index of 2.6 shows a large
amount of small scale clumps and filaments, whereas the fBm
with $\beta=3.7$ consists basically of one peak with fragmented
boundaries.

\begin{figure}
\centering
\epsfig{file=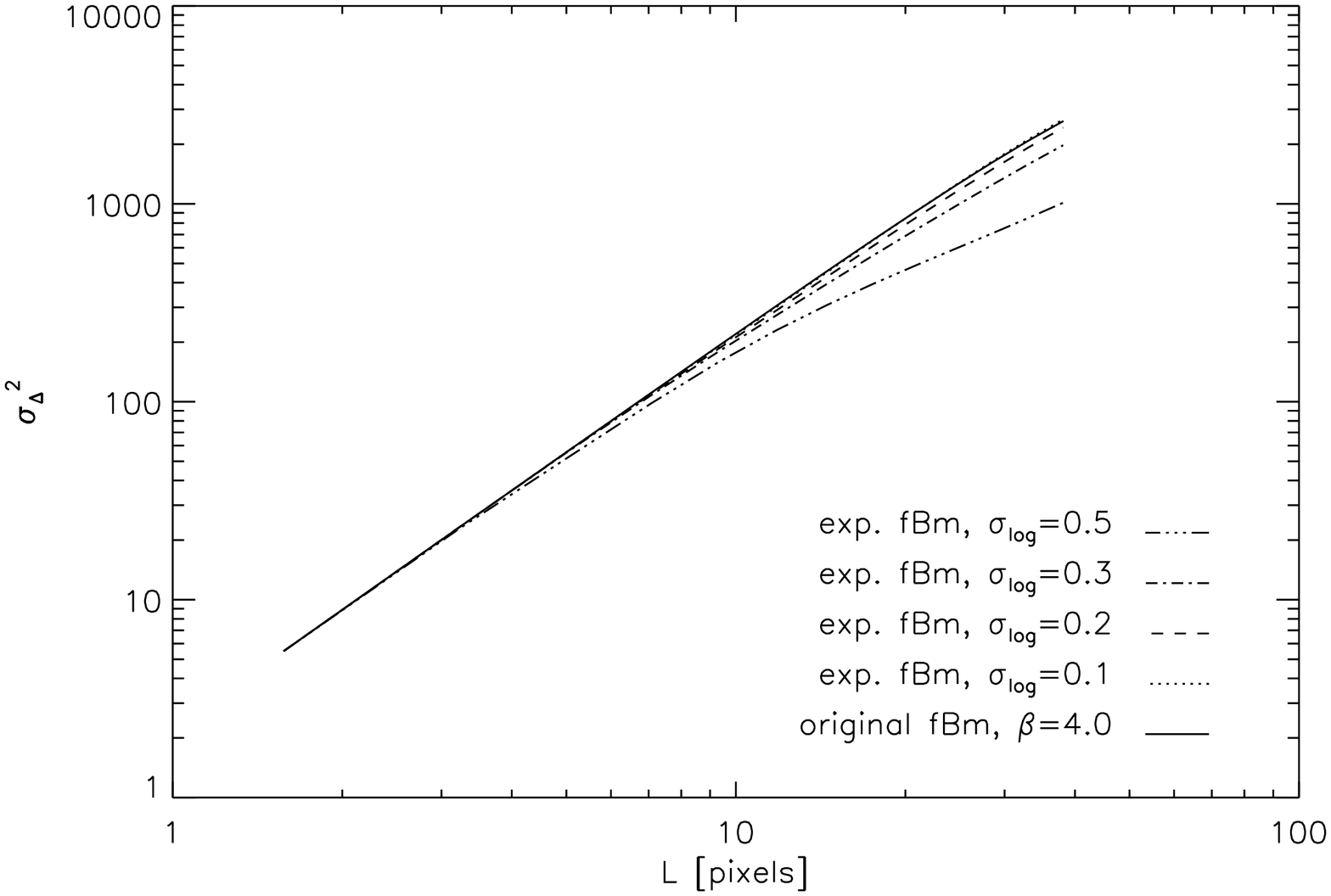,angle=90,width=\columnwidth}\\
\epsfig{file=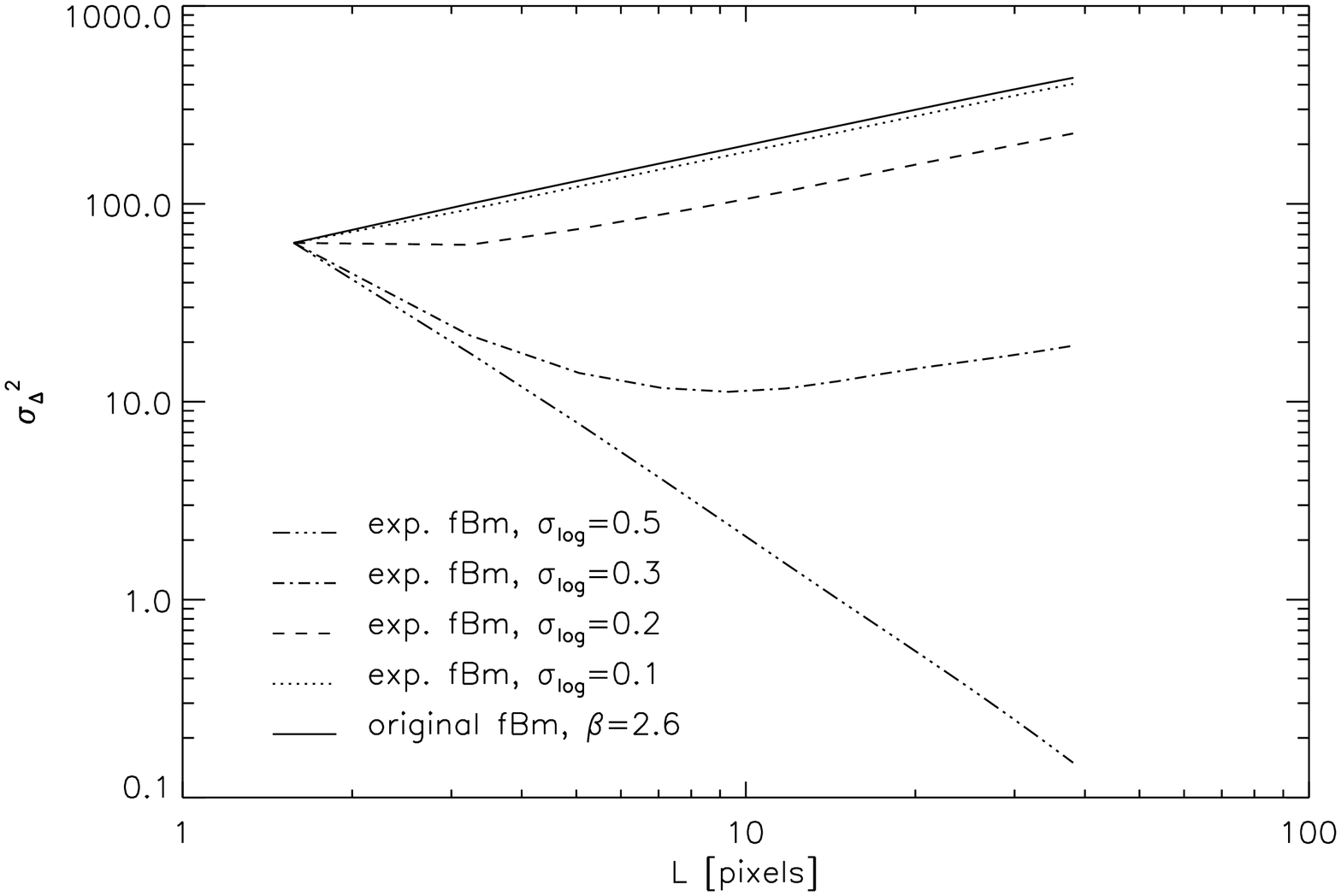,angle=90,width=\columnwidth}
\caption{$\Delta$-variance spectra of the projected structure of 
exponentiated fBm's compared to the spectrum of the original fBm.
The upper plot represents $\beta=4$, the lower plot $\beta=2.6$.
The different lines indicate different stretching factors $a$
resulting in different logarithmic widths of the distributions.
The average logarithmic density is taken to be 2.0 in all cases.}
\label{fig_exponentiate}
\end{figure}

The figure also reveals a general problem of fBm's when interpreted
as density structure. They show negative values. fBm's have
on the average a Gaussian probability distribution with vanishing
mean so that negative values can only be avoided when adding a large
constant density offset. However, in this way we drastically
change the ratio $\sigma_{\rho}/\rho_0$ for the data set. 
Another method to create a density distribution containing only
positive values is to square or exponentiate the original fBm
as proposed by \citet{Stutzki98}. {\changed \citet{Miville} have
claimed that exponentiation, $\rho\sub{exp}=\rho_0 \exp(a \rho\sub{fBm})$,
does not affect the power-spectrum,
but it is mathematically obvious, that it can potentially destroy
the power-law scaling. Thus we have tested the impact of exponentiation
for different spectral indices $\beta$ and different factors $a$
translating the standard deviation of the fBm into the logarithmic
standard deviation of the new density structure. The result is shown
in Fig. \ref{fig_exponentiate} for an fBm with $\beta=4$
as used in Fig. 11 of \citet{Miville} and for an fBm with $\beta=2.6$
in terms of $\Delta$-variance spectra (see Sect. \ref{sect_deltavar}).
It is obvious that for narrow distributions, the distortion of the
original spectrum by exponentiation is small, as the exponentiation
is then close to a linear transformation. In general, we have to
acknowledge, however, considerable distortions of the spectrum by
the exponentiation. When creating a very wide density distribution
from the $\beta=2.6$ fBm we
even find a completely different scaling behaviour resembling
rather a structure with $\beta=0$. The example from 
\citet{Miville} corresponds approximately to the $\beta=4$,
$\sigma\sub{log}=0.3$ case shown in Fig. \ref{fig_exponentiate}.
Here, the deviation from the original spectrum is small so that
it was not detectable. {\newchanged Moreover, we have found that
the $\Delta$-variance reacts much more sensitive to the exponentiation
than the azimuthally averaged power spectra. Only for very wide
distributions and low spectral indices the azimuthally averaged power
spectra show similar noticeable deviations. In general we have to
conclude that exponentiation results in a change of 
the scaling properties.} Consequently, non-linear transformations 
are not well suited to produce well defined test data for the
density structure.} We will stick to the simple approach
of adding a constant to the fBm and ignoring the remaining 
negative values for the construction of the density structure.
The implications of this approach are quantified in detail in
Sect. \ref{sect_zerolevel}. In contrast to the density structure
which has to be positive defined and necessarily a non zero mean
$\rho_0$, the velocity structure can
directly use fBm's thus guaranteeing a zero value for $v_0$ so
that the simplifications discussed above apply.

When using {\newchanged independent} fBm's to represent both the density
and the velocity
structure of interstellar clouds we neglect, however, the
interrelation of both quantities in the interstellar medium
determined by the hydrodynamic equations, especially by the Poisson
equation. Comparisons with magneto-hydrodynamic simulations by \citet{Esquivel} have
shown, however, that the {\newchanged cross-correlation between the density 
and velocity fields has} a negligible effect
on the centroid velocities so that we can neglect {\newchanged its impact} here.
We will {\newchanged further discuss} the influence of cross-correlations between
density
and velocity structure on different observational parameters
in a subsequent paper.

\subsection{The $\Delta$-variance}
\label{sect_deltavar}

The $\Delta$-variance analysis was introduced by \citet{Stutzki98} 
and improved and extended by \citet{Bensch} and \citet{OKS}. Here, we
repeat only those definitions which are essential for
the centroid analysis.

The $\Delta$-variance in a structure $f(\vec{x})$ is computed
by filtering the data set with a spherically symmetric normalised
wavelet of characteristic size $l$, consisting of a positive
inner part and a negative annulus, and computing the variance of the 
filtered map. \citet{OKS} have tested various wavelet shapes, but their
mutual differences are not significant for the analysis performed here
so that we stick to the ordinary French hat filter from \citet{Stutzki98}.
The $\Delta$-variance is then the variance of the filtered map, as a
function of the filter size, given by
\begin{equation}
\sigma_\Delta^2(l)= \left\langle \left( f(\vec{x}) * {\bigodot}_l(\vec{x}) \right)^2 \right\rangle_{\vec{x}}
\end{equation}
where the symbol $*$ stands for a convolution, $\bigodot_l$ describes
the filter wavelet and the average is taken over the whole data set.
If $l$ is the average distance between two points in the core and the
annulus in the filter, the $\Delta$-variance spectrum $\sigma_\Delta^2(l)$
measures the amount of structure on the given scale $l$. 

The $\Delta$-variance is related to the power spectrum of a structure
$P(\vec{k})$ by
\begin{equation}
\sigma_\Delta^2(l)= \int P(\vec{k})\; 
\left| \tilde{\bigodot}_l(|\vec{k}|) \right|^2 \; d^n\vec{k}
\label{eq_deltafourier}
\end{equation}
where $\tilde{\bigodot}_l$ is the Fourier transform of the
filter function with the size $l$ and $\vec{k}$
denotes the spatial frequency or wavenumber. In case of
isotropic structures the power spectrum is spherically
symmetric, $P(\vec{k})=P(|\vec{k}|)$. This is also the case
for the Fourier transformed filter function as long as it is
spherically symmetric in the spatial domain.
The power spectrum is given by the Fourier transform of the
autocorrelation function 
\begin{equation}
A(\vec{l})=\left\langle f(\vec{x}) f(\vec{x}+\vec{l}) \right\rangle_{\vec{r}}
\label{eq_acf}
\end{equation}

For power-law power spectra \citet{Stutzki98} showed 
that for 2-D structures in the interval of spectral indices $\beta$
between 0 and 6, the $\Delta$-variance spectrum is as well
a power law with the exponent $\alpha=\beta-2$. In three
dimensions the range is extended to $0 < \beta < 7$ and the
exponent is $\alpha=\beta-3$. Equivalent slopes are obtained
locally in case of non-power-law power spectra. However,
in this case there is no analytic relation for the normalisation
factor of the $\Delta$-variance spectrum so that it can only
be obtained by numeric integration.

Thus the $\Delta$-variance is basically a very robust method
to evaluate the power spectrum of a structure.  The advantages of the
$\Delta$-variance compared to the direct computation of the power
spectrum result from the smooth filter shape which provides
a very robust way for an angular average independent from gridding
effects, and from the insensitivity to edge effects as discussed
by \citet{Bensch}. {\changed A possible disadvantage is the implicit
radial averaging, which does not allow to search for signatures
of anisotropy still contained in the two-dimensional power spectrum
$P(\vec{k})$. Such an anisotropy was considered by \citet{Esquivel}
but is irrelevant for our studies.}

\subsection{Comparing $\Delta$-variance and structure function}
\label{sect_compare_delta_sf}

\citet{Lazarian03} and \citet{Esquivel} used the (second order)
structure function instead of the $\Delta$-variance to characterise the
scaling of velocity centroids. The structure function is related as well to
the autocorrelation function, $D(\vec{l})=2\left[ A(0)-A(\vec{l}) \right]$,
\citep[see e.g.][]{Miesch94}. With the power spectrum being the
Fourier transform of the autocorrelation
function we also have a trivial relation between structure functions
and power spectra.

For structures with a power-law power spectrum \citet{Stutzki98}
studied analytically the relation between the power spectrum, the 
autocorrelation function and the $\Delta$-variance. They find in the range 
of spectral indices $3< \beta < 5$ in 3-D and for $2 < \beta < 4$
in 2-D, and in the limit of infinitely large data sets,
power-law structure functions. {\changed Using the notation of \citet{LP00},
this is the range of steep spectra. Here, the spectral index of the
structure function} agrees with the index of the $\Delta$-variance
spectra discussed above.
In the range of shallow spectra with lower power spectral indices,
$0 < \beta < 3$ or $0 < \beta < 2$ respectively,
the autocorrelation function {\changed is  a power 
law\footnote{For spatial separations
corresponding to wavenumbers {\changed smaller} than the cut-off wavenumber 
given by the finite sampling of any system.}
so that the structure function must deviate from a power law
behaviour.
The structure function is always increasing with lag towards
the maximum} given by twice the total variance of the structure
$\sigma_f^2=A(0)$. 

For MHD simulations producing basically steep velocity spectra
but with significant deviations from pure power laws \citet{OML}
compared the centroid velocity structure function
with the $\Delta$-variance of the centroid map and showed that both
give a similar scaling behaviour, having comparable slopes within
a large part of the spectrum., The $\Delta$-variance, however, is
advantageous with respect to the detection of pronounced scales in the map
and is more robust with respect to observational artifacts.
Altogether, the $\Delta$-variance seems to be somewhat better suited 
to determine the exponent of the power spectrum, as it shows itself
a larger range of a power-law behaviour and it is more stable with
respect to observational restrictions.

On the other hand, \citet{Esquivel} demonstrated that the
structure function of centroid velocities can be analytically
understood with respect to its composition from density and
velocity fluctuations. This represents a clear advantage
relative to the $\Delta$-variance. Thus we have actually performed
all tests of the centroid structures reported here both with the
$\Delta$-variance analysis and with the structure
function. As a surprising result we find very little differences
in the general behaviour. Therefore, we concentrate in the following
analysis on the $\Delta$-variance spectra discussing the differences
in comparison to the structure functions only in Sect. \ref{sect_sf_2}.

\subsection{Projection effects}
\label{sect_projection}

The relation between a 3-D structure and projected
2-D maps, obtained by the integration along the line
of sight, has been studied in detail both in terms of the
$\Delta$-variance \citep[e.g.][]{Stutzki98, MLO} and of
the structure function \citep{Esquivel}. A projection
of the density structure $\rho(\vec{x},z)$ is inherently
performed when observing the intensity map $I\sub{int}(\vec{x})$ of
an optically thin tracer in a medium of constant excitation
temperature (Eq. \ref{eq_zconvol}).

The projection effect on the $\Delta$-variance spectrum can be easily
understood by realizing that the $\Delta$-variance is basically a
robust method to deduce the power spectrum. In Fourier space, projection 
corresponds to the selection of the zero-frequency component in the
considered direction. For isotropic structures the power spectral indices of
projected maps in any direction agree with the spectral index
of the 3-D structure. This is fulfilled by definition for
the fBm structures used here for testing. Thus the local slope of
the power spectrum $\beta$
is retained, and all components with non-zero spatial frequencies
in the considered direction are dropped. Because the $\Delta$-variance
is obtained by convolving this power spectrum with the Fourier transform of
either a 3-D or a 2-D wavelet, the resulting spectrum
has a local slope $\alpha\sub{3D}=\beta-3$ or $\alpha\sub{2D}=\beta-2$,
respectively. The mutual translation is straight forward. {\newchanged
The exponent of the power spectrum is retained on projection, while the
index of the $\Delta$-variance\footnote{The same applies to the structure
function, but in a limited spectral range.} is increased by one.}
This has been confirmed in the application of the $\Delta$-variance
analysis to the 3-D density structure of (magneto-)hydrodynamic simulations
and their projection onto maps by \citet{MLO}.

For power-law power spectra, the translation of the amplitudes can also
be performed analytically following the formalism provided in the Appendix of
\citet{Stutzki98}. As an approximation we can also use the simple empirical
relation
\begin{equation}
\sigma_{\Delta,{\rm 3D}}^2(l) = \sigma_{\Delta,{\rm 2D}}^2(l) \times \frac{l}{l\sub{cube}}
\times 1.97 \exp\left(-\frac{\beta}{2.83}\right)
\label{eq_deltaproj}
\end{equation}
which is accurate within a few percent for power spectral indices $\beta$
between 1 and 4 and cube sizes of at least $32^3$ pixels. Even for sufficiently
smooth, but non-power-law $\Delta$-variance spectra Eq. (\ref{eq_deltaproj})
can be applied by using an index $\beta(l)$ derived from the local
slope.

A general problem is, however, the actual loss of information
by projection. There is no way to recover the Fourier amplitudes which
are dropped by the projection. Thus the re-translation from the 2-D
$\Delta$-variance spectrum into the corresponding 3-D spectrum
is only possible by assuming isotropy. \citet{MLO} and \citet{OML} studied the
degree of anisotropy in hydrodynamic and magneto-hydrodynamic simulations
by comparing 2-D and 3-D $\Delta$-variance spectra and found
that the assumption is clearly violated for simulations with strong magnetic
fields but reasonably justified for most other simulations.

\begin{figure}
\centering
\epsfig{file=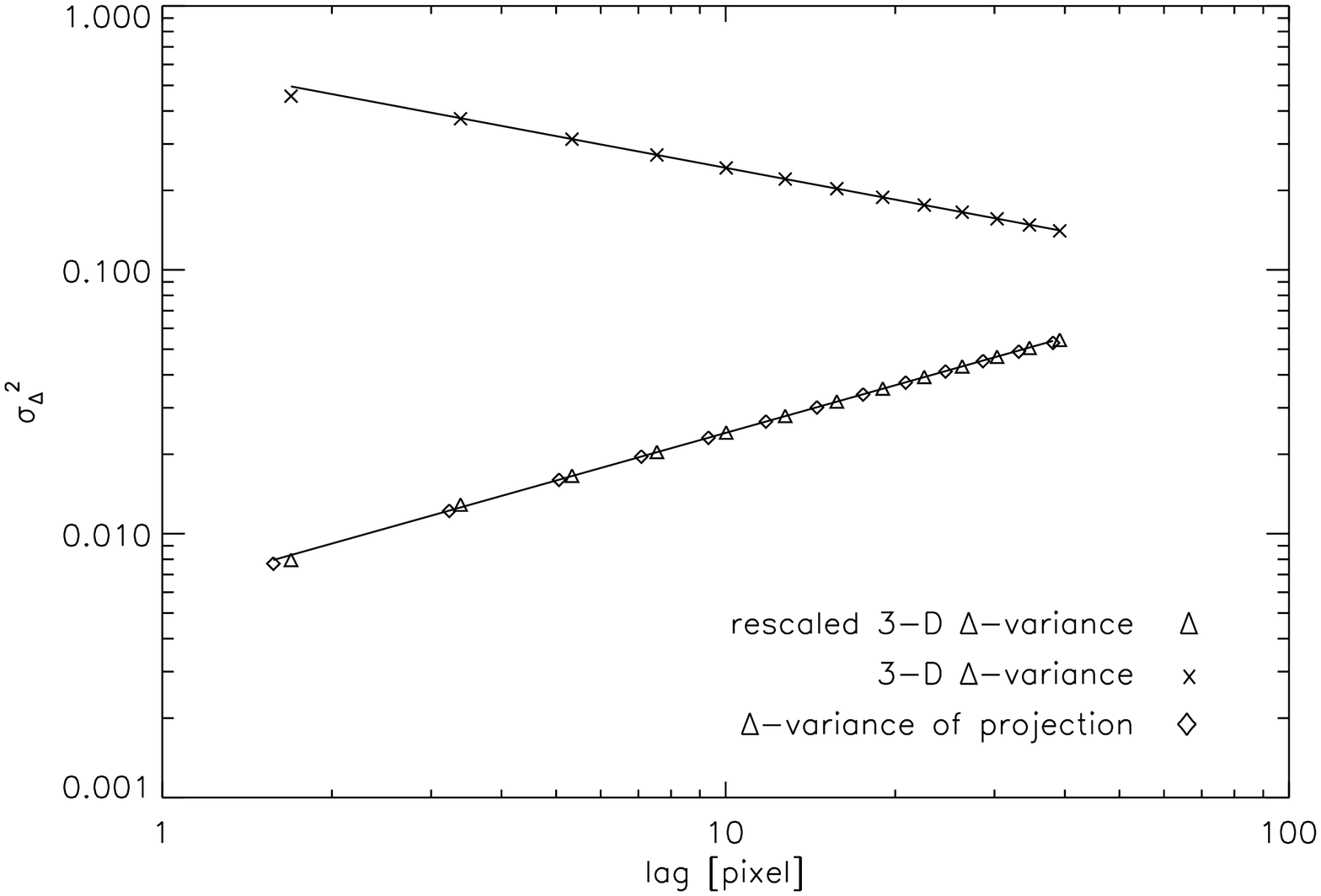,angle=90,width=\columnwidth}\\
\epsfig{file=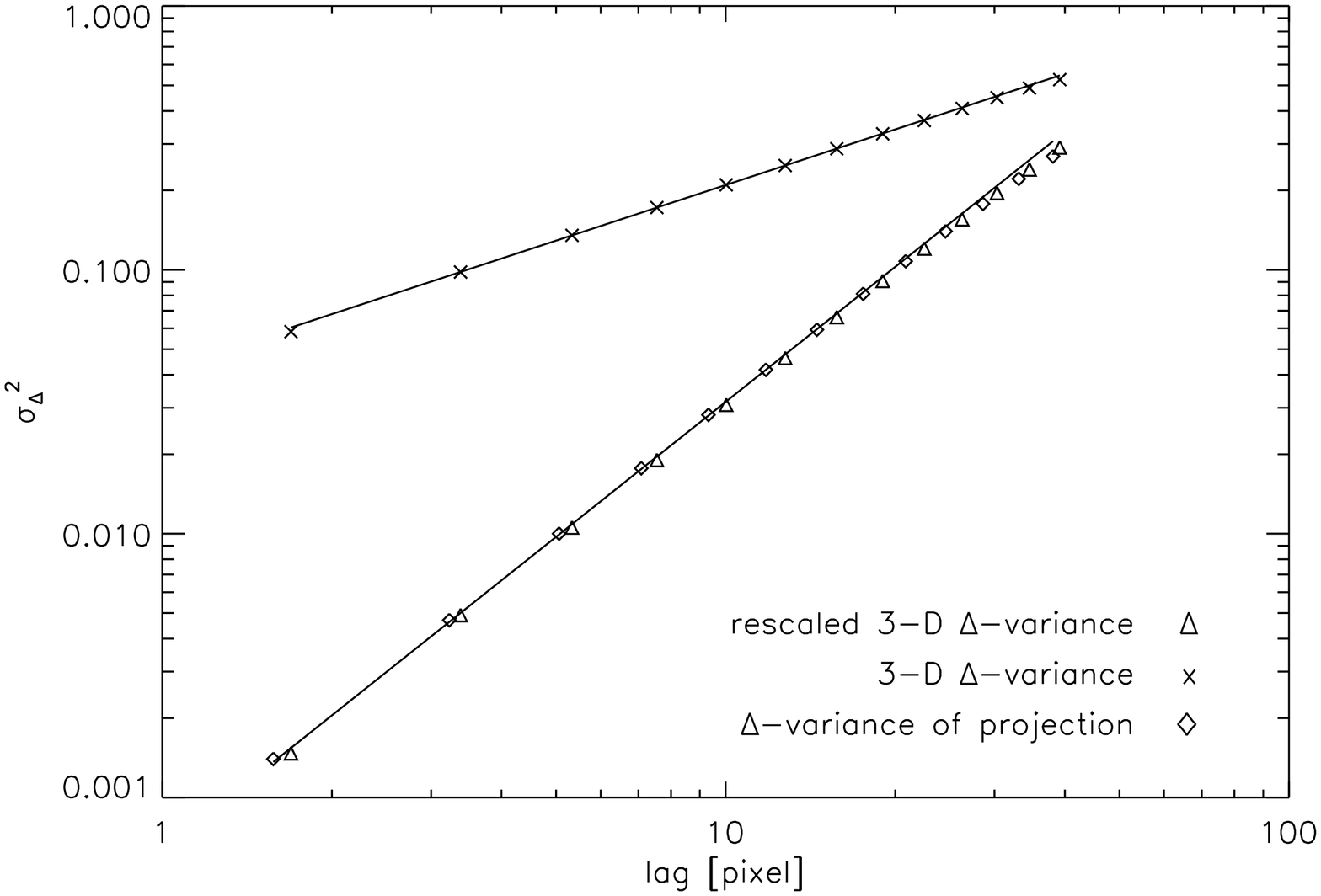,angle=90,width=\columnwidth}
\caption{$\Delta$-variance spectra determined in
3-D structures and their projection together with the translation
of the 3-D $\Delta$-variance spectrum into the corresponding
2-D spectrum using Eq. (\ref{eq_deltaproj}).
For the upper plot an fBm with $\beta=2.6$ shifted by $2\sigma$ and 
truncated at zero was used. The lower plot shows the result for an
fBm with $\beta=3.7$. The solid lines mark slopes corresponding to
the spectral indices $\beta-3$ and $\beta-2$.}
\label{fig_projections}
\end{figure}

Fig. \ref{fig_projections} demonstrates the influence of the projection
effects on the $\Delta$-variance spectra of two fBm's. The upper graph
represents an fBm structure with a shallow index $\beta=2.6$ and the
lower graph a steep spectrum with $\beta=3.7$. The $\Delta$-variance
spectra measured in 3-D and for the projected structure follow
almost exactly the theoretical power-law relation with the exponents
$\alpha=\beta-3$ or $\alpha=\beta-2$, respectively. The triangles
stand for the results from the $\Delta$-variance computed in 3-D and
translated into a 2-D spectrum using Eq. (\ref{eq_deltaproj}). We
find an excellent agreement with the spectra obtained directly
from the projected maps.

Beyond the plotted range the $\Delta$-variance spectra show a
turn-over at about {\newchanged half} of the total size of the simulated
cube arising from the lack of larger structures due to the periodicity condition
in the construction of the data \citep[see][]{Bensch}. Because of the lacking
significance at large lags, the spectra are only computed
up to lags of about a third of the cube size. 

\section{Centroid composition effects}

Taking their relation to the autocorrelation function
both the $\Delta$-variance and the structure function of velocity
centroid maps will be given by
averages of the products $v\sub{c}(\vec{x}) v\sub{c}(\vec{x}+\vec{l})$
(see Eq. \ref{eq_acf}).  Using the decomposition
of the velocity centroids in Eq. (\ref{eq_centroids}) and assuming a zero
average velocity $v_0$ we see that four terms characterise the scaling:
\begin{eqnarray}
&&\!\!\!\!A_{v\sub{c}}(\vec{l})=\rho_0^2 \left\langle
\int dz\; \delta v(\vec{x},z) \times
 \int dz\; \delta v(\vec{x}+\vec{l},z) \right\rangle_{\vec{x}}
\nonumber \\
&&+
\left\langle
\int dz\; \delta \rho(\vec{x},z) \delta v(\vec{x},z) \times
\int dz\; \delta \rho(\vec{x}+\vec{l},z) \delta v(\vec{x}+\vec{l},z)  \right\rangle_{\vec{x}}
\nonumber \\
&&+ 
\rho_0 \left\langle
\int dz\; \delta v(\vec{x},z) \times
\int dz\; \delta \rho(\vec{x}+\vec{l},z) \delta v(\vec{x}+\vec{l},z)  \right\rangle_{\vec{x}}
\nonumber \\
&&+
\rho_0 \left\langle
\int dz\; \delta \rho(\vec{x},z) \delta v(\vec{x},z) \times
\int dz\; \delta v(\vec{x}+\vec{l},z) \right\rangle_{\vec{x}}
\label{eq_general_decomposition}
\end{eqnarray}
The first term is the autocorrelation function of the projected velocity
fluctuations.
If this term dominates, the scaling behaviour of the centroid velocities 
reflects exactly the scaling behaviour of the velocity structure. In this case
it is easy to deduce the properties of the velocity structure from an observed
map of centroids. We find the simple projection of the velocity structure
onto a 2-D map like in the case of the column density map reflecting the
3-D density structure.
The second term {\newchanged describes a}  combination of the fluctuations of the density
and the velocity structure. The term also contains the
mutual correlation between density and velocity fluctuations along the
line of sight. The third and fourth terms 
quantify the cross-correlation between velocity fluctuations at one
point and density fluctuations at another point. In case of isotropic media
both terms are identical. {\changed They should statistically vanish
in case of {\newchanged independent} density and velocity structures, but some
remainders due to accidental {\newchanged cross-correlations} are expected for any
particular realisation.} A similar decomposition in terms of the
structure function was provided by \citet{Lazarian03}.
From the decomposition in Eq. (\ref{eq_general_decomposition}) we see that the
{\changed ratio} between the average density and the density fluctuations
should provide a criterion whether the centroid map is a good
measure for the scaling of the velocity field. Thus we test in the following
the composition of centroid velocity maps from fBm structures adjusting
their parameters in such a way that {\changed the full range of observed
spectral indices for the density and velocity structure in the
interstellar medium is} covered.

\subsection{The density zero level}
\label{sect_zerolevel}

A major problem with the artificial simulation of density structures
is the mutual incompatibility of Gaussian fluctuations and strictly
positive values for the density. As discussed in Sect. \ref{sect_fbms}, fBm structures
always show a Gaussian distribution of values and the analytic expressions
for the velocity centroids derived by \citet{Lazarian03} are also
based on the assumption of Gaussian fluctuations. {\newchanged 
However, as long as the average of a Gaussian distribution is not large
compared its dispersion, negative values are unavoidable for sufficiently
large samples.}

A common way to create positive densities is to add a constant
density until the minimum value in the cube falls at zero \citep{Miville,
Esquivel03}. A major drawback of this method is, however, that
the minimum value of a Gaussian distribution depends on the
exact realization of the random numbers used to generate the distribution
and it is very sensitive to the size of the data cube. Thus
the added value, then providing the average density $\rho_0$,
may significantly vary from simulation to simulation. By renormalising
the average density to {\newchanged unity} as proposed by \citet{Miville}
and \citet{Esquivel03} the variation is only transferred
to the standard deviation of the density distribution because
the ratio between standard deviation and mean is retained. Moreover,
the approach results typically in $\sigma_{\rho}/\langle \rho
\rangle < 0.3$ \citep{Miville}. Such values are in {\changed
contradiction to many observational data \citep[see e.g.][]{Jenkins}.
Density fluctuations with $\delta \rho/\rho \approx 1$ are expected for
Mach numbers approaching unity. Such Mach numbers characterise
warm media, while colder parts of the ISM tend to have supersonic
velocities \citep[see][]{Elmegreen} leading to
even larger density fluctuations \citep{Falgarone98,Padoan}.}  .

\begin{figure}
\centering
\epsfig{file=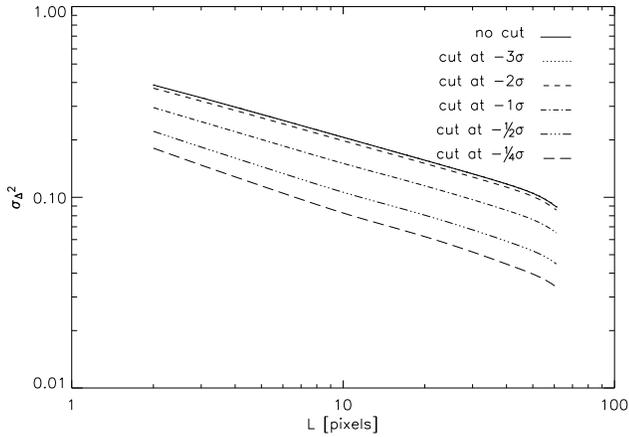,angle=90,width=\columnwidth}
\caption{$\Delta$-variance spectra of 3-D density structures
obtained by shift-and-truncate from an fBm with $\beta=2.6$
and different truncation levels. {\changed The spectrum for 
the density cut at $3\sigma$ is partially indistinguishable
from the original spectrum.}}
\label{fig_density_truncation}
\end{figure}

To a certain extent these problems can be circumvented by {\changed
combining the density shift with a truncation 
of the residual negative tail.
When we shift the density distribution e.g. by $\rho_0=1 \sigma_{\rho}$,
by adding this constant value, and discard all points falling below
zero} only 8\,\% of the points {\changed from the original distribution
are set to a zero value} so that the statistical properties of the 
overall structure are hardly changed.
In this way we can obtain positive densities and a
$\sigma_{\rho}/\langle \rho \rangle$ ratio of about one, avoiding all problems
from a dependency on the resolution and on the exact random numbers. One has to
keep in mind, however, that the truncation of the density structure 
{\newchanged can} have a noticeable influence on the scaling properties.
{\changed The
pure addition of the constant density does not affect affect the spectrum
because it is scale-independent.}

To test the possible error {\changed introduced by the truncation 
of the distribution at a given density level} we have analysed the
truncated fBm's and compared them to the original spectra. The result is shown
for a spectral index $\beta=2.6$ and different truncation levels
in Fig. \ref{fig_density_truncation}. The actual shift of the
density by $\rho_0$ does not influence these spectra {\changed because
the $\Delta$-variance is insensitive to any constant offset. We
see the pure truncation effect.} For truncation levels {\changed of 
$2\sigma_{\rho}$ and above} the spectra
are practically not changed. For truncation levels between
$0.5 \sigma$ and $2\sigma$ the shape of the spectra is retained
but they are shifted to lower absolute values. This can be explained
by the reduction of the total variance in the data cubes {\changed which
is visible in the scale-dependent $\Delta$-variance as well.} 
{\changed The original distribution was normalised to a variance of unity
in this example whereas the truncation leads to reduced 
variances of 0.96, 0.76, and 0.56 for the $2\sigma$, $1\sigma$, and $0.5\sigma$
truncation levels}, respectively. These {\changed are exactly the
numbers by which the $\Delta$-variance spectra in Fig. \ref{fig_density_truncation}
are shifted relative to the original spectrum.} Only for a truncation level
at {\changed $1/4\sigma$ the slope of the spectrum is changed,
i.e. the scaling behaviour of the structure is modified. In this case
the absolute shift of the $\Delta$-variance spectrum also does no longer
exactly match the corresponding reduction of the total variance of the 
density distribution relative to the original value.} 

Examining the resulting projected maps shows that the relation between
the 3-D scaling and the 2-D scaling given in Eq. (\ref{eq_deltaproj})
is also preserved down to truncation levels of $0.5\sigma$.
Corresponding studies for different spectral indices show that
the $\Delta$-variance spectra are least sensitive to truncations
at low spectral indices, {\changed between 2 and 2.5,} where even truncation levels of 
$0.25\sigma$ do not change the scaling behaviour and the
relation between total variance and $\Delta$-variance.
At spectral indices close to four, in contrast, the
$0.5\sigma$ truncation plot shows already significant deviations,
so that we conclude that a negligible statistical impact on the
scaling behaviour is only guaranteed at truncation levels around
$1\sigma$ and above. {\newchanged The shift-and-truncate method to
create positive densities is thus not perfect in terms of retaining
the original scaling properties of the structure, but the 
introduced deviations are still small compared to those introduced
by the non-linear transformations discussed in Sect. \ref{sect_fbms}.
They} would be hardly detectable
in observed data, although we have to take them into account when
performing a detailed quantitative analysis.

\begin{figure}
\centering
\epsfig{file=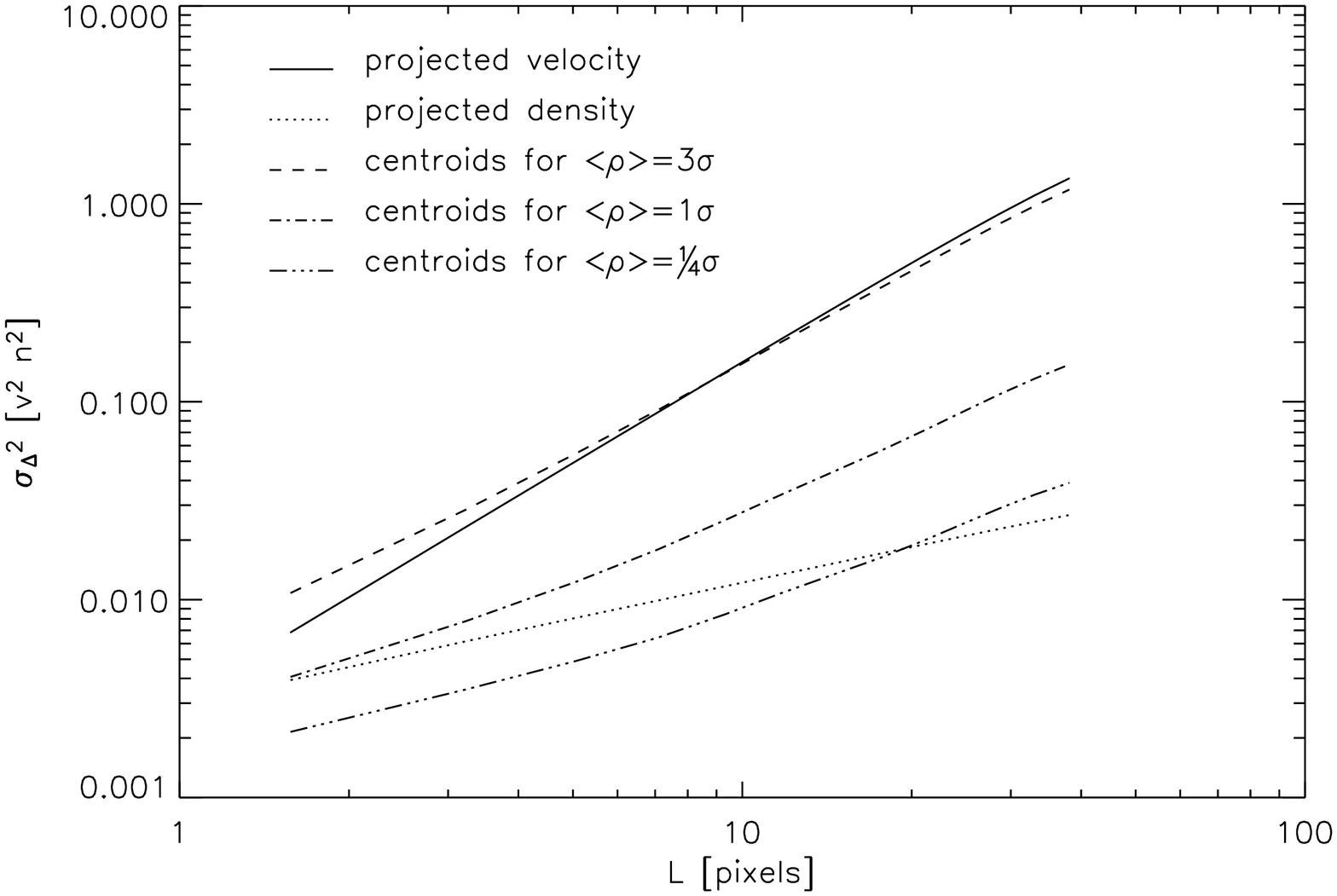,angle=90,width=\columnwidth}\\
\epsfig{file=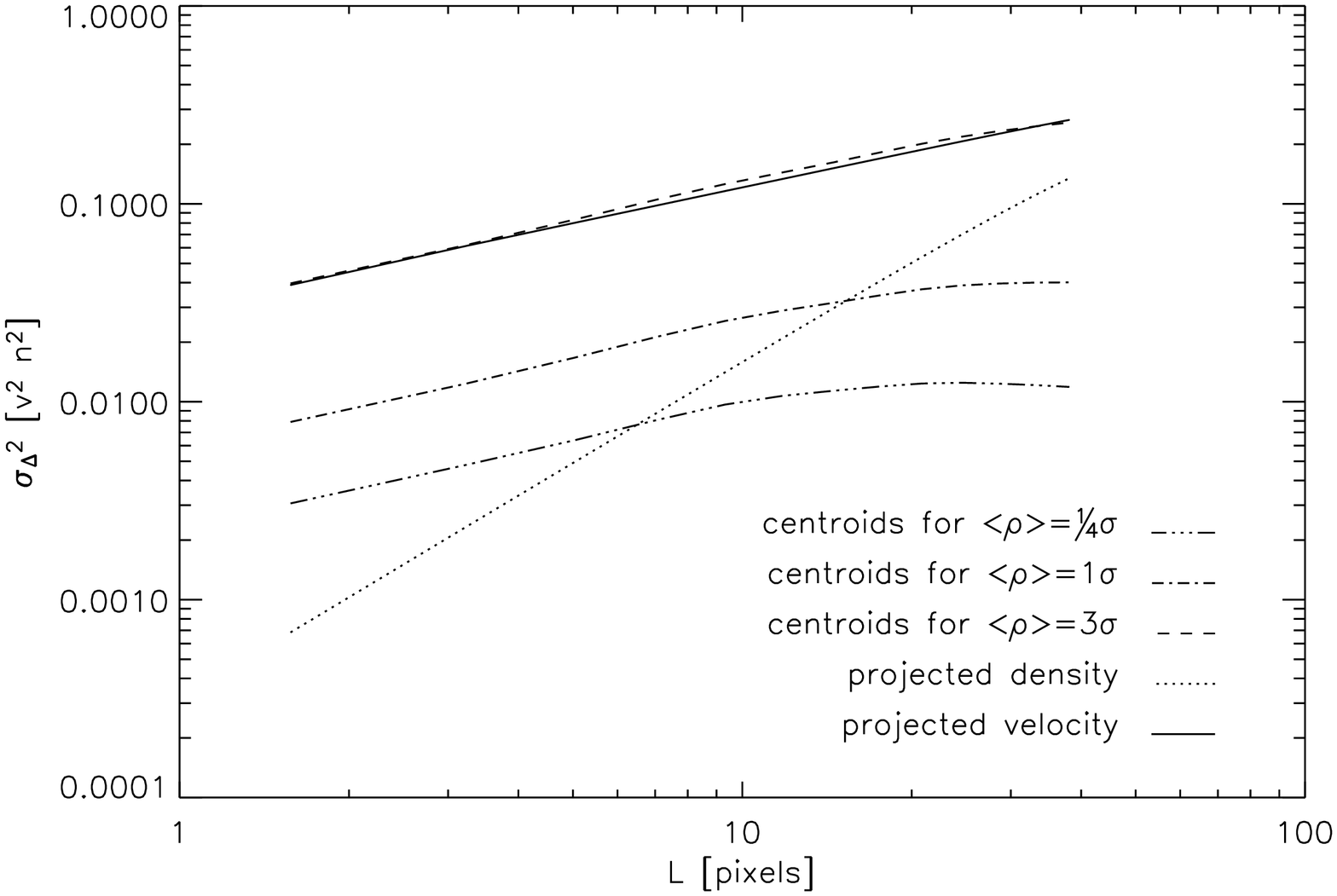,angle=90,width=\columnwidth}
\caption{Comparison of the $\Delta$-variance spectra of the
weighted centroid velocities with the spectra of the original
density and velocity structure for different truncation levels. 
To plot equivalent quantities the projected density is multiplied by
$\langle v^2 \rangle$ and the projected velocity by $\langle
\rho^2 \rangle$ \citep[see][]{Esquivel}. The $\langle \rho^2 \rangle$
factor was computed for the $3\sigma$-cut density cube. The
corresponding plots for $1\sigma$ and $0.25\sigma$ would be shifted
down by a factor 5.2 and 13.5, respectively. The upper plot
was computed from an fBm density structure with
$\beta=2.6$ and a velocity structure with $\beta=3.7$, the lower
plot used the opposite spectral indices.}
\label{fig_delta_densitytruncation}
\end{figure}

Fig. \ref{fig_delta_densitytruncation} shows two actual examples
for the influence of the density zero level definition on the
measured centroid velocity spectra. The scaling behaviour of the
centroids was computed in terms of the $\Delta$-variance spectra
for three different shift-and-truncate levels of fBm generated
density structures. To judge how far they reflect the original
density or velocity structure, we have also plotted the $\Delta$-variance
spectra of these projected quantities multiplied with the mean square
of the complementary quantity to guarantee units equivalent 
to the centroids. 

The upper plot shows the combination of a shallow density spectrum with a
steep velocity spectrum, matching a situation which is typically observed
in {\changed molecular clouds (see Sect. \ref{sect_fbms}). The absolute shift of the
curves for weighted centroids is mainly determined by the different
values of $\langle \rho^2 \rangle$ produced by different average
densities. However, this shift does not influence the characteristic
scaling behaviour within the structure. Looking at the slopes of the
centroid spectra, we} find a confirmation of the general
considerations on the role of the density zero level $\rho_0$ given
above. If the density structure is dominated by a large average, i.e.
in the case of $\rho_0=3\sigma_{\rho}$,  the centroid velocities
are basically given by a projection of the velocity structure,
so that they reproduce the original velocity scaling behaviour.
For lower average densities, i.e. a lower relative contribution
of the pure velocity projection given by the first term in Eq.
(\ref{eq_general_decomposition}), the centroid scaling becomes shallower
with an exponent which is close to that of the velocity structure
at large scales and an exponent which is close to that of the density 
structure at very small scales and the lowest values of $\rho_0$.
This plot seems to confirm the transition from purely velocity-dominated
centroids to density-dominated centroids as originally interpreted
by \citet{Lazarian03}.

If we consider, however, the opposite situation of spectral
indices in the lower plot, we only find that the centroid
scaling becomes less and less representative for the actual
velocity structure when reducing the average density $\rho_0$.
Their scaling does not tend towards the scaling
of the {\newchanged column} density structure but becomes shallower as well. This fact
is confirmed in all simulations with other combinations of
spectral indices. At low values of $\sigma_{\rho}/\rho_0$, the
centroids match the projected velocity structure, whereas
their scaling becomes shallower for lower average densities
irrespective of the actual spectral index of the density
structure. The spectral index of the density structure
determines, however, at which scales the deviations occur.
For density structures with a shallow spectral index, dominated
by many small-scale fluctuations, the main effect occurs at
small scales. In contrast, we find the main deviations at
large scales, when the density structure has a steep index,
representing a relative dominance of large scale fluctuations.
When interpreting changes in the slope of the $\Delta$-variance
spectrum of observed centroid maps we can thus use the known
information on the projected density scaling to judge whether
they represent an actual deviation of the velocity structure
from self-similarity or whether they might be produced just by
the centroid composition effects. In general, we find a bigger
impact on the overall centroid spectrum as shallower both 
spectra become. However, is not clear that the a shallow velocity field is
physically motivated \citep{Esquivel}.

\begin{figure}
\centering
\epsfig{file=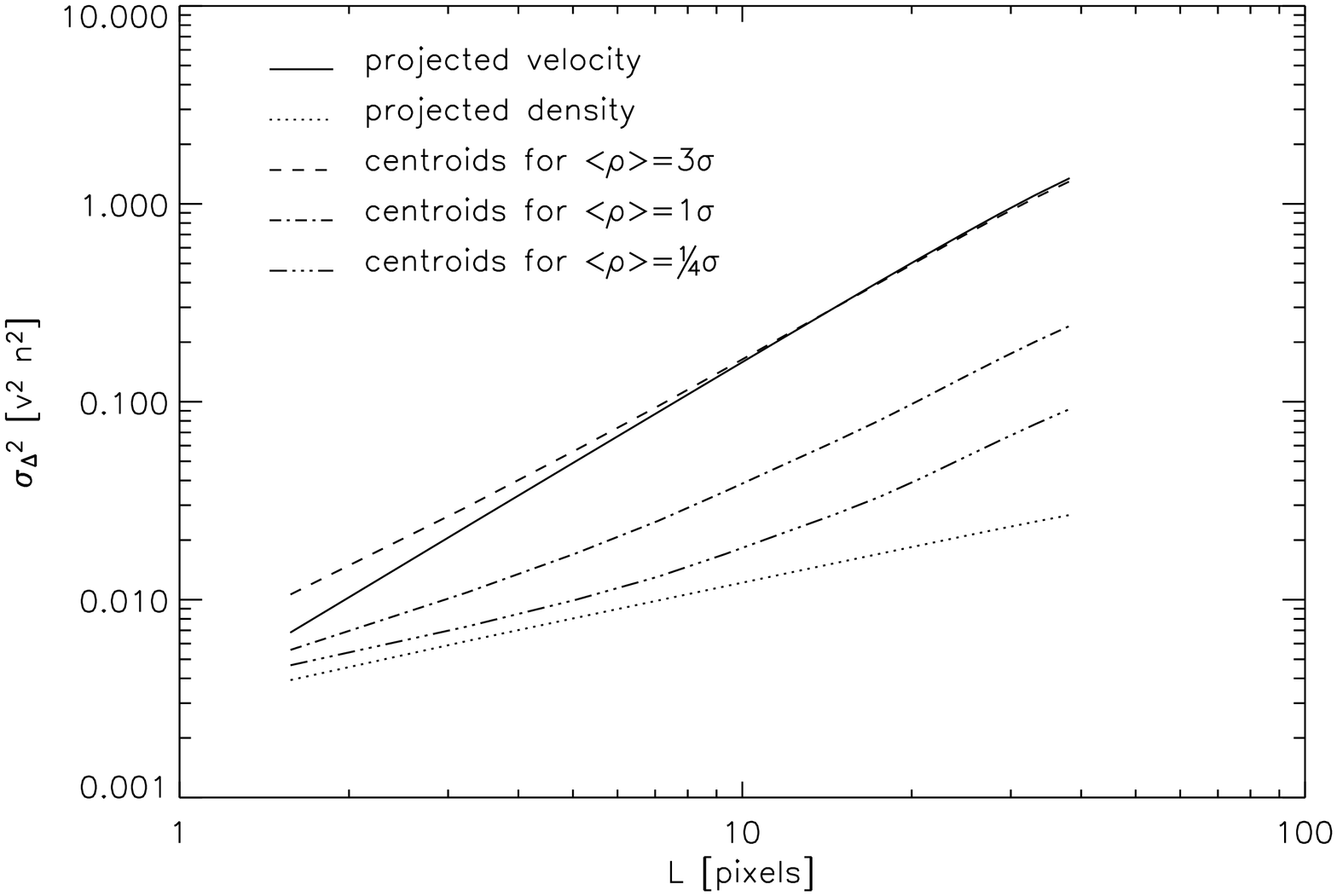,angle=90,width=\columnwidth}\\
\epsfig{file=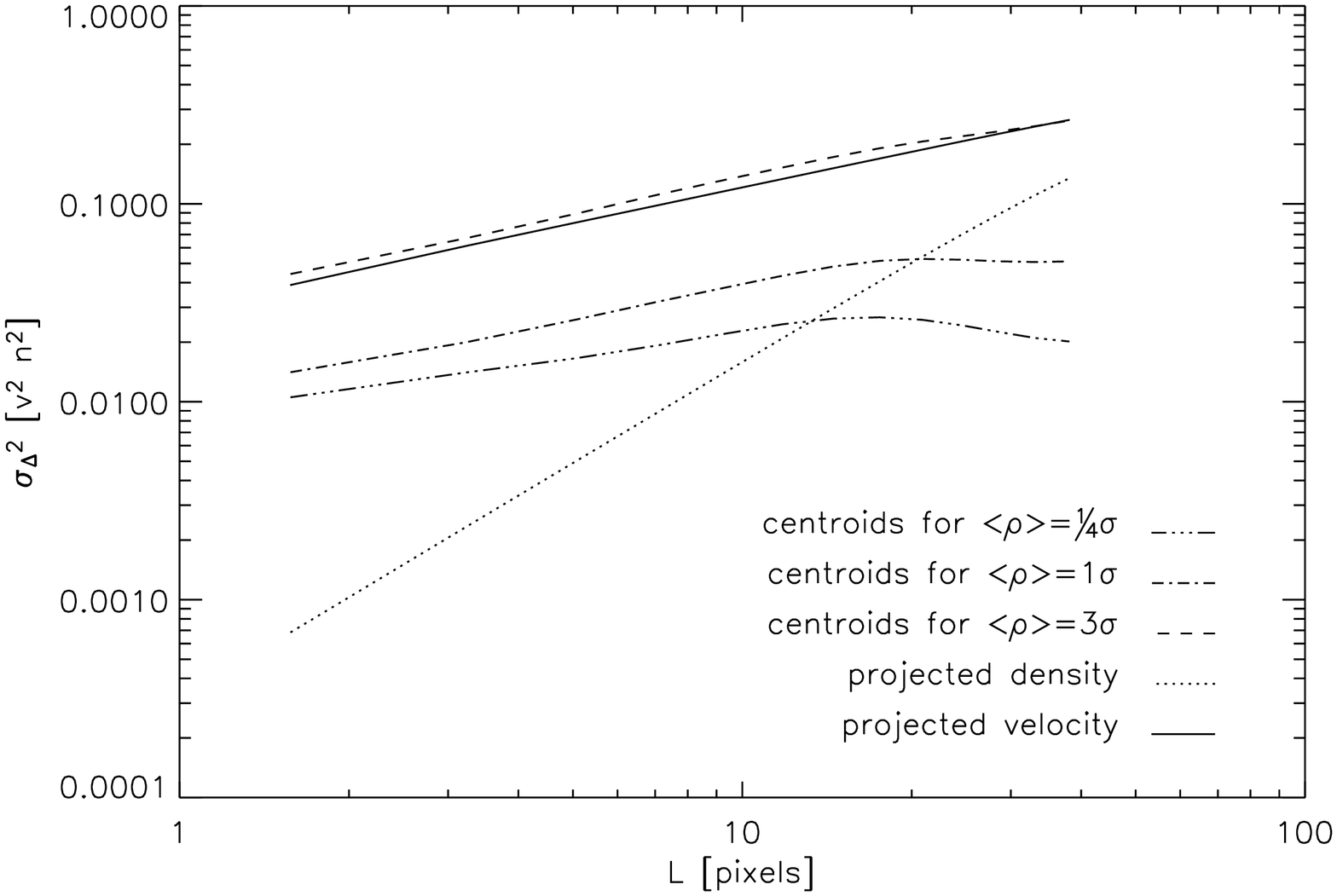,angle=90,width=\columnwidth}
\caption{Same as Fig. \ref{fig_delta_densitytruncation} but
for normalised centroids. They are rescaled by the factor $\langle \rho^2
\rangle$ to obtain comparable dimensions.}
\label{fig_delta-normal_densitytruncation}
\end{figure}

In Fig. \ref{fig_delta-normal_densitytruncation} we show the same
effects for the ordinary, normalised centroids instead of the
weighted centroids used in Fig. \ref{fig_delta_densitytruncation}.
We find the same general behaviour as for the weighted centroids
but differences in details. In all cases with high average
densities, i.e. for $\sigma_{\rho}/\rho_0<1$ the normalised
centroids provide a slightly better reproduction of the
original velocity structure than the weighted centroids.
At lower densities, they are somewhat weaker modified for shallow
density spectra and somewhat stronger modified for steep density
spectra. The modifications correspond approximately to the
same effect that a change of $\sigma_{\rho}/\rho_0$ by a factor
1.5 would have for the weighted centroids. In general we can state, however, that either
both centroid definitions reveal the true velocity structure
or none of them. The direct retrieval of the velocity scaling from
the $\Delta$-variance spectra of the centroids will only succeed
when the average cloud density is significantly larger than
the density dispersion. In these cases the normalised centroids
are marginally better than the weighted centroids.

These results explain the differences and agreements between
the previous studies on velocity centroids discussed so far in the literature.
The studies of \citet{Miville} and \citet{Lazarian03} used a relatively
large average density and they indicated a good match between centroid spectra
and projected velocity spectra. The mechanism of producing positive 
densities from fBm's by adding a large constant used by
\citet{Miville} and \citet{Esquivel03} gave results {\changed
that correspond to our results for applying the 
shift-and-truncate technique with a large average density, i.e. when
we add the $3\sigma$ density offset.} Both centroid definitions
follow the actual velocity scaling over a large range of
scales, deviating at most at the very ends of the spectra
in this case. One has to emphasise that this matching is only 
produced by adding a large $\rho_0$ value, so that the general
conclusion that centroids are a good measure for the
velocity structure drawn by \citet{Miville} and applied to interpret
observational data by \citet{Miville2} 
does not hold {\changed for the general case of interstellar gas
with substantial density fluctuations.}

In cases with lower average densities, all using
a combination of steep velocity spectra with shallow density
spectra,  \citet{OML}, \citet{Lazarian03}, and \citet{Brunt}
found centroid spectra which were shallower than 
the velocity spectrum. The hydrodynamic and magneto-hydrodynamic
turbulence models studied \citet{OML} were characterised by
steep velocity spectra with $\beta_v\approx 4.0$, 
shallow density spectra with $\beta_{\rho} \approx 2.5 \dots 2.7$,
and a high density contrast with $\sigma_{\rho}/\rho_0 >5$
thus corresponding closely to the conditions for the
low density curve in the upper plot of Fig. \ref{fig_delta-normal_densitytruncation}.
%
With a limited dynamic range for fitting the $\Delta$-variance
spectra it is obvious that the flattening of the centroid
spectra relative to the original velocity spectrum seen in
the figure can be misinterpreted as a constant reduction of the slope by
one.


Our results can also explain the findings of \citet{Brunt} studying
the characteristics of velocity centroids of HD and MHD turbulence
simulations as a function of Mach number. {\changed With the known
relation between Mach number and density dispersion \citep{Padoan},
their finding of a growing discrepancy between the average spectral
index of the velocity distribution and of the centroid map with
growing Mach number} can be explained by the impact of an increasing
$\sigma_{\rho}/\rho_0$ ratio, which reduces the relative
contribution of the projection term. This is most clearly
seen in the models of decaying turbulence where, e.g., for an
initial $\sigma_{\rho}/\rho_0$ ratio of 1.1 the centroid spectrum is
shallower by 0.8 than the velocity spectrum whereas it is for the
final $\sigma_{\rho}/\rho_0$ ratio of 0.5 only shallower by 0.1.
We have to emphasise, however, that this approach cannot explain the
differences in the spectral indices obtained by \citet{Brunt} for MHD models
observed perpendicular or parallel to the main magnetic field
direction. In these cases, the isotropy assumption used in our
analysis is clearly violated.

\subsection{The velocity zero level}

{\newchanged The composition of weighted centroids (Eq. \ref{eq_centroids})
is a priori symmetric with respect to density and velocity. 
{\changed In the decomposition in Eq. (\ref{eq_general_decomposition}), 
we have assumed, however,
that the velocity scale is chosen in such way that $v_0=0$ while
$\rho_0>0$}. To better understand the centroid behaviour it is useful
to perform an experiment using velocity fields with $v_0>0$. From
the symmetry of the problem, we expect that we find a centroid behaviour
matching the density scaling for large average velocities $v_0$
in the same way as we find centroids matching the velocity
scaling for large average densities $\rho_0$. Indeed, we obtain
the new term 
$ v_0^2 \left\langle \int dz\; \delta \rho(\vec{x}) \times
 \int dz\; \delta \rho(\vec{x}+\vec{l}) \right\rangle_{\vec{x}}$
in Eq. (\ref{eq_general_decomposition}) if $v_0\ne0$. It
contains the spectrum of the projected density fluctuations.
In contrast to the density treatment, we do not apply any truncation 
to the velocity structure when shifting it to $v_0>0$. The experiment
thus provides an additional test for the significance of the truncation.
If the simple shift of the velocity structure behaves equivalent
to the shift-and-truncate of the density structure we can be sure that
all effects result from the selection of the average values and not
from the truncation.}

\begin{figure}
\centering
\epsfig{file=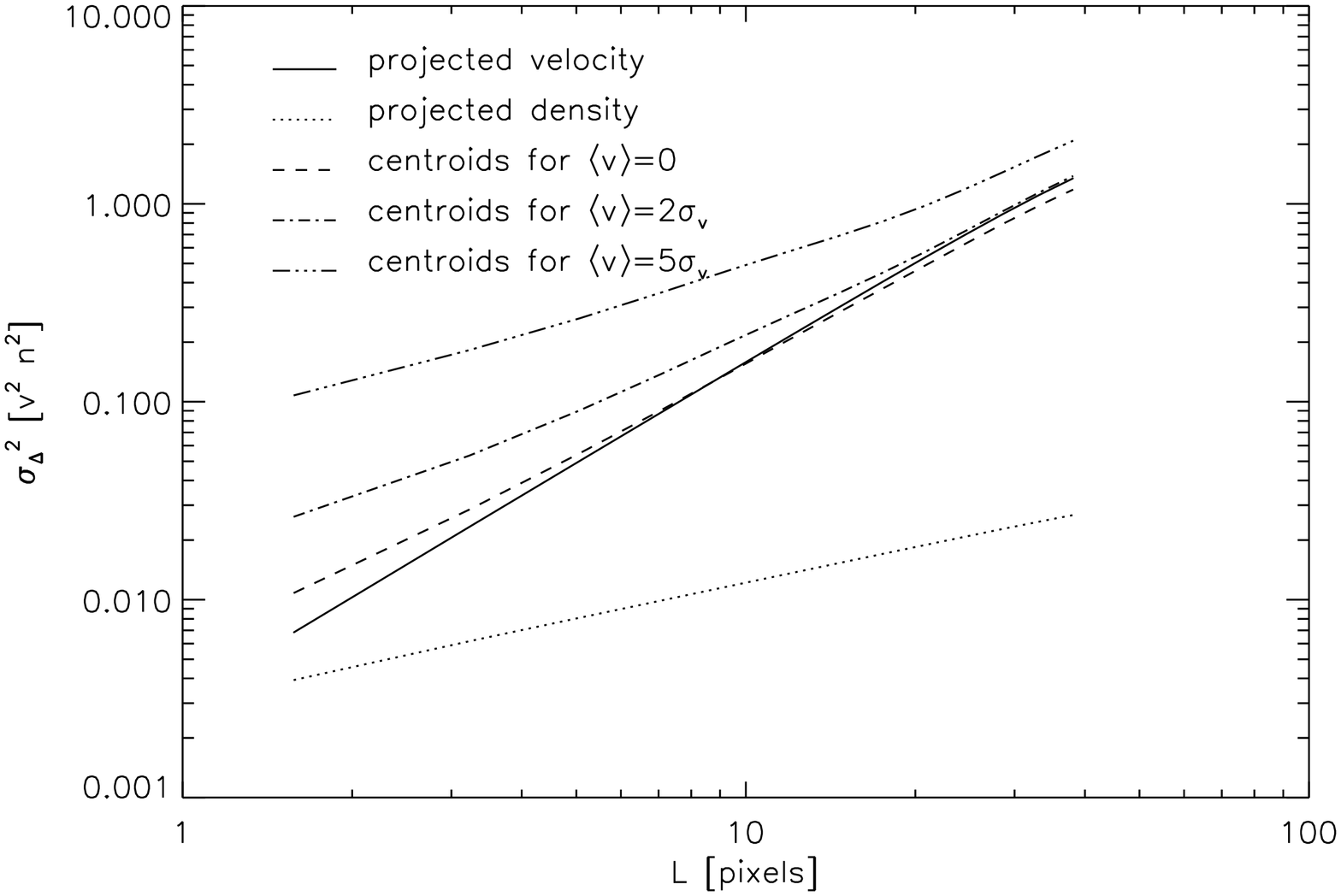,angle=90,width=\columnwidth}\\
\epsfig{file=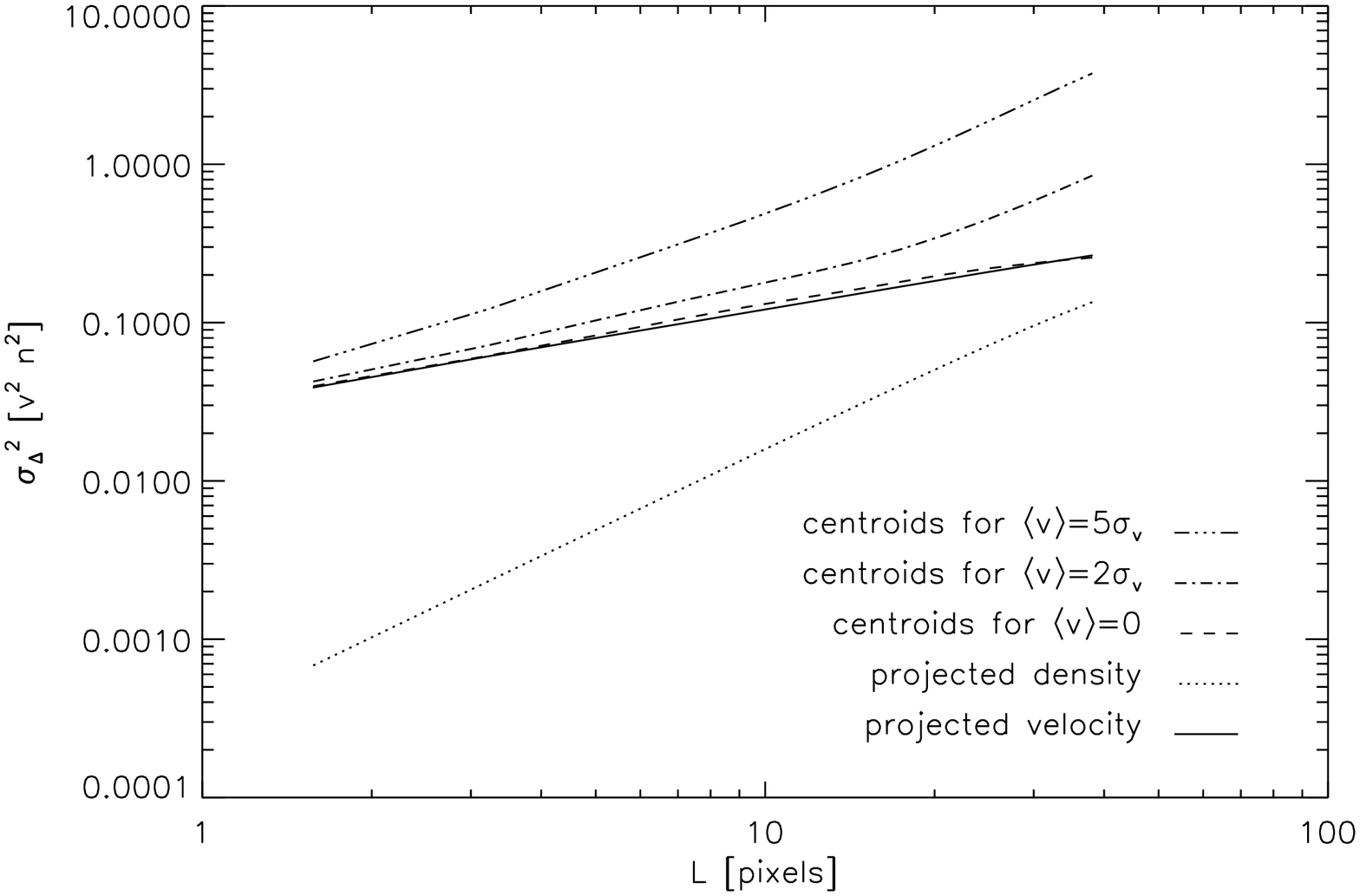,angle=90,width=\columnwidth}
\caption{$\Delta$-variance spectra of the centroid velocities computed using
the density and velocity structures from Fig. \ref{fig_delta_densitytruncation}
and different $v_0$ levels. {\changed A density structure
with $\langle \rho \rangle = 3\sigma_{\rho}$ was used here.}}
\label{fig_delta-v0}
\end{figure}

Fig. \ref{fig_delta-v0} shows the impact of different velocity
offsets on the centroid $\Delta$-variance spectra. A high average 
density, {\changed $\rho_0=3\sigma_{\rho}$ was chosen,} to guarantee that the
centroid spectrum for $v_0=0$ is dominated by the velocity
structure. {\changed The curves for $v_0=0$ are identical to the
$\rho_0=3\sigma_{\rho}$ curves in Fig. \ref{fig_delta_densitytruncation}.}
 When increasing the average velocity we find in the
upper plot a transition to shallower spectra similar to the
effect of a reduced average density in Fig. \ref{fig_delta_densitytruncation}.
The slope of the centroid spectrum remains close to the
slope of the velocity spectrum at large scales and at
small scales it takes the slope of the {\newchanged column} density spectrum. {\changed
In the lower panel we find as well that the centroid spectrum is more
and more similar to the {\newchanged column} density spectrum when increasing the average
velocity. This is {\newchanged opposite to the effect of reducing the}
average density in the lower panel of Fig.
\ref{fig_delta_densitytruncation}.}
{\changed The adjustment} of the average velocity reproduces the
transition from velocity-dominated spectra to density-dominated
spectra, {\newchanged as predicted by \citet{Lazarian03}.

The equivalence of the impact of the velocity shift on the
centroid spectra to the impact of the shift-and-truncate method for
the density structures proves that the main change of the
centroid spectrum is due to the added offsets and not due to} the 
truncation of the density structure at its low density wing.
{\changed Unfortunately, the {\newchanged numerical experiment
 cannot be exploited to derive} the
true velocity scaling when the average density is so small that the
centroid spectra for $v_0=0$ are ``density-contaminated''.
By {\newchanged increasing $v_0$ we will} only increase the contribution from the
density scaling, which is already known from the projected intensity
maps, but we cannot remove the effect of the combination of
density and velocity fluctuations.}

\subsection{Further decomposition}

\begin{figure}
\centering
\epsfig{file=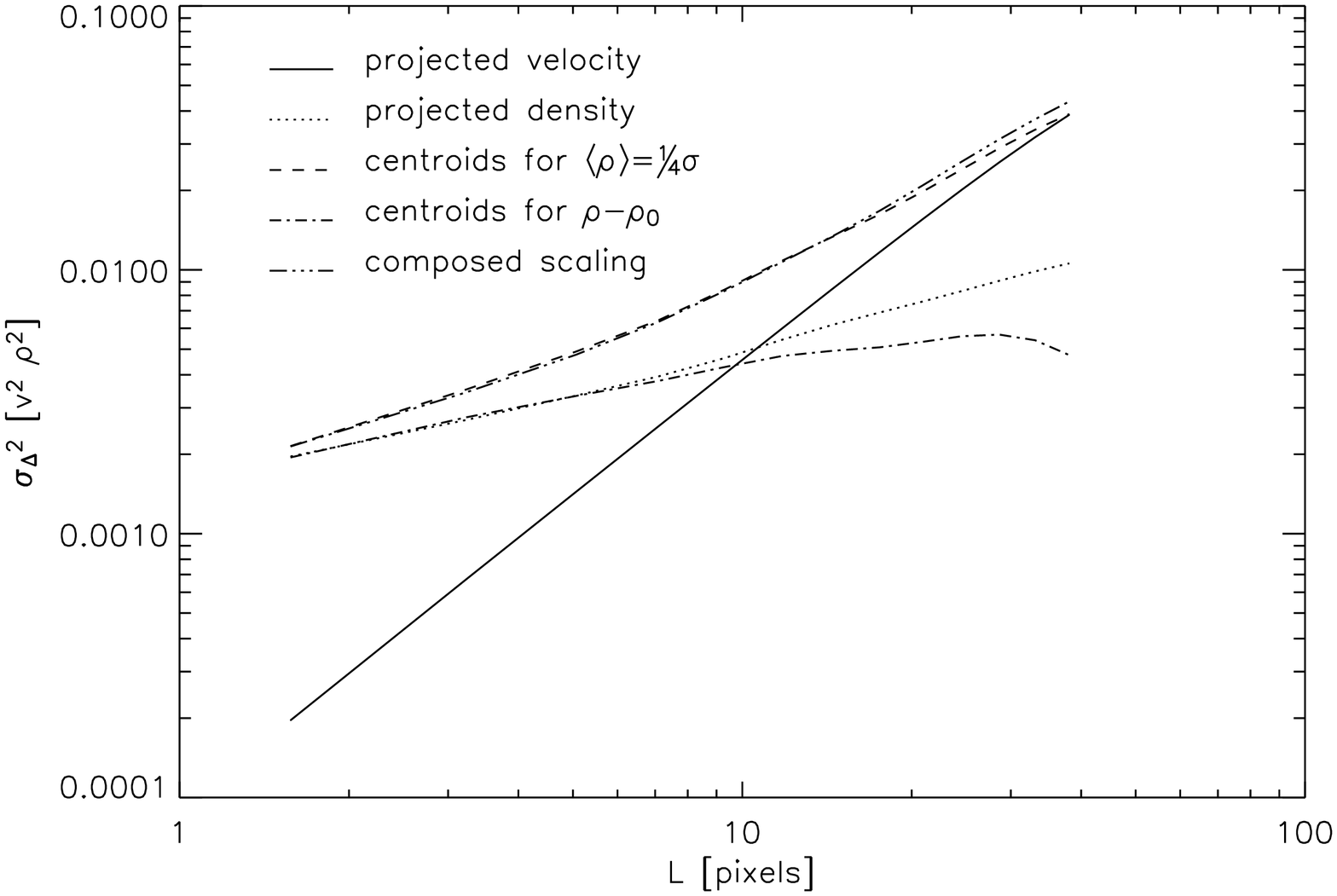,angle=90,width=\columnwidth}\\
\epsfig{file=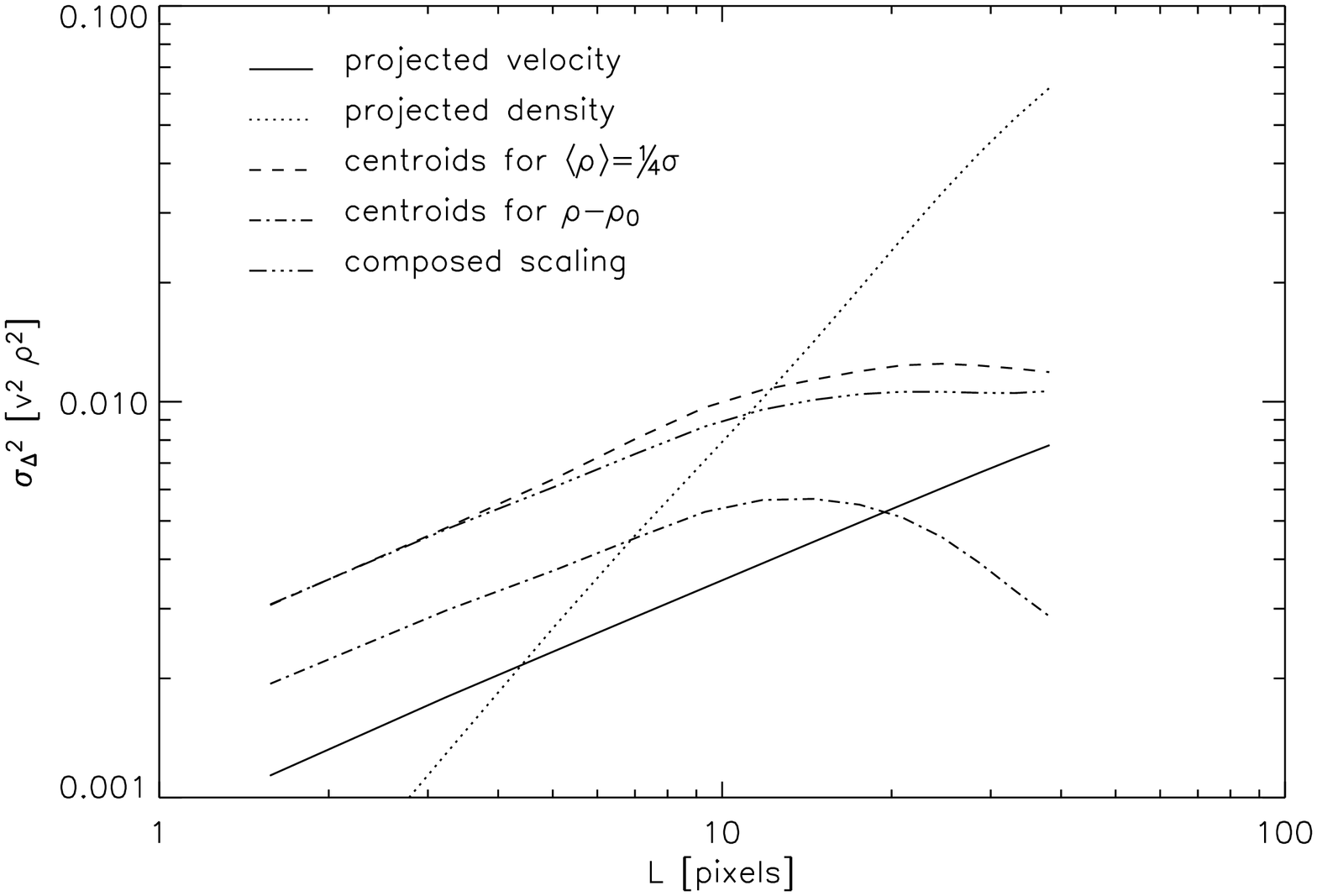,angle=90,width=\columnwidth}
\caption{Decomposition of the $\Delta$-variance spectra of the
weighted centroid velocities for the density and velocity structures
from Fig.  \ref{fig_delta_densitytruncation} using shift-and-truncate by
1/4 of the original standard deviation for the density structure.
The solid line represents the contribution from the velocity structure
projected with a $\rho_0$ weighting and the dash-dot line represents
the centroid contribution from the {\changed auxiliary} density
fluctuation field. The sum
of both terms (dash-dot-dot-dot) is very close to the measured centroid
spectrum (dashed).
}
\label{fig_delta-decomposition}
\end{figure}

The results obtained so far show that the density zero level
basically changes the contribution from the first term in Eq.
(\ref{eq_general_decomposition}) representing the pure
projection of the velocity structure. In a next step we 
investigate the relative contribution of the other three terms
to the deviation measured between the projected velocity spectra
and the centroid spectra.
{\newchanged The third and fourth term vanish if there is no
cross-correlation between the density and velocity fields.
For our independently generated fBm structures this should be
the case.  We expect, however, that in every realization some
accidental correlations occur so that the two terms
are only negligible in the ensemble average.}

{\newchanged If the density field is known, we can obtain the second term,
i.e. the combination of density and velocity fluctuations,
by constructing an auxiliary density field $\rho\sub{aux}=\rho-\rho_0$
and computing the weighted centroids for this auxiliary quantity.
As the average density of the auxiliary field vanishes, the derived
centroids directly match the second term in Eq. (\ref{eq_general_decomposition}). 
This procedure is illustrated in Fig. \ref{fig_delta-decomposition} 
where we plot the centroid spectrum for the auxiliary
field $\rho\sub{aux}$, and compare the full centroid spectrum obtained
from the original density structure with the sum of this second term 
and the pure velocity scaling term.}
For the sake of comparison we also plot the spectrum of the
projected density and velocity structure, where the velocity
spectrum is multiplied here by $\rho_0^2$ to represent
exactly the first term in Eq. (\ref{eq_general_decomposition}).
The same combination
of spectral indices as used in Fig. \ref{fig_delta_densitytruncation}
was taken. A $0.25\sigma_{\rho}$ shift-and-truncate level was used for
the density structure, so that the centroid spectrum deviates
considerably from the spectrum of the velocity fluctuations.

For all studied combinations of spectral indices, we obtain a
good match between the sum of the projected velocity spectrum
and the centroid spectrum from the {\changed auxiliary field of density
fluctuations with the full spectrum of the velocity centroids. 
{\newchanged Nevertheless, we find always a non-negligible
difference between the two curves, resulting from the
accidental cross-correlations contributing to the third and 
fourth term, which are not contained in the sum.}
 We also find that the second
term, giving the combination of all fluctuations, has a spectrum
which is always shallower than either of the projected spectra
involved. For a steep spectrum of density
fluctuations its slope turns even negative at large scales. 
This explains why the total spectrum of the weighted centroids
is always shallower than the projected velocity spectrum,}
independent of the spectral index of the density spectrum.

{\changed The computations have confirmed the theoretical
expectation, that the spectrum of velocity centroids consists
of only two main contributions:} the pure projection of the velocity
structure determined by the average density $\rho_0$ and a
shallow term mainly determined by the density fluctuations.
In Sect. \ref{sect_iterate} we show how this
decomposition can be exploited to measure the actual velocity structure
from observed centroid maps if the $\sigma_{\rho}/\rho_0$ ratio
can be estimated independently.

\subsection{Matching criteria}

We have seen that the single quantity giving the ratio between
the strength of the density fluctuations and the average density
$\sigma_{\rho}/\rho_0$ is able to discriminate between the different
behaviours of the centroid spectra. For low values of this ratio,
the spectra are dominated by the actual velocity structure so that
the 3-D velocity scaling is preserved in observed centroid maps.
For $\sigma_{\rho}/\rho_0\la 0.5$ the
$\Delta$-variance spectrum of the centroid map directly measures 
the spectral index of the underlying velocity structure.
For higher values, the centroid spectra are always shallower than
the spectra from the projected velocity maps.
They are produced by a combination of density and velocity fluctuations.

In contrast to the suggestion of a density-dominated regime by
\citet{Lazarian03}, the systematic study of a wide range of combinations
of spectral indices with the $\Delta$-variance spectra shows no
indications of a transition from velocity dominated
centroids to density-dominated centroids, but rather a transition to 
"density-contaminated" spectra. Using a decomposition of structure
functions similar to Eq. (\ref{eq_general_decomposition}) \citet{Lazarian03}
identified a term that indeed traces density fluctuations. 
{\newchanged They showed that in general centroids 
do not trace directly the velocity fluctuations.}
However, in their numerical tests, they use a combination of steep velocity
and shallow density spectra, and disregard a cross term that is
equivalent to the convolution of velocity and density fluctuations
presented here. {\changed The shallow centroid spectrum was interpreted
as the density spectrum. We have demonstrated} that for the $\Delta$-variance spectra
a density-dominated regime arises only if we chose a velocity scale 
with an offset so that $v_0\ne 0$. {\changed However, the combination
of the facts that the centroid spectrum is always shallower than the
velocity spectrum and that most observed density spectra are shallower
as well} can give the false impression that
centroids trace the density scaling for large ratios $\sigma_{\rho}\rho_0$.

{\changed \citet{Esquivel} presented another criterion for a match between
centroid and velocity scalings, namely,}
$X^2 \sigma_{v\sub{c}}^2\gg \langle v^2 \rangle \sigma_{I\sub{int}}^2$.
{\changed They stated, however, that it is not clear how large the ratio
$X^2\sigma_{v\sub{c}}^2/(\langle v^2 \rangle \sigma_{I\sub{int}}^2)$ 
should eventually be to guarantee} that the centroids reliably represent
the velocity structure.
When applied to the overall data cubes we find that the distinctive
power of the criterion is limited.
In the examples plotted above we obtain {\changed for instance a ratio
of 35 when using the $3\sigma$ shift-and-truncate level of the density
distribution and a ratio of 3.0 for the $0.25\sigma$ shift-and-truncate
level} in the case of the shallow density and steep velocity spectrum.
{\changed In contrast, we obtain corresponding ratios of 2.6 and 0.32,
respectively, for the combination of steep density and shallow velocity
spectrum.}
 In both cases the {\changed $3\sigma$ shift-and-truncate level} 
gives a good match between centroid and velocity scaling while
the $0.25\sigma$ level results in a very poor agreement. {\newchanged
Thus the global criterion is poorly quantified.}

{\changed The criterion can be rewritten in a scale-dependent form
\citep{Lazarian03}: $X^2 D_{v\sub{c}}(\vec{l})/(\langle v^2 \rangle 
D_{I\sub{int}}(\vec{l})) \gg 1$, when we  consider the
structure function of the two maps at a given lag $l$.
One} might assume that this criterion should
hold as well for $\Delta$-variance spectra because of their
similar scaling properties. Then a ratio
$X^2\sigma_{\Delta,v\sub{c}}^2(l)/(\langle v^2 \rangle 
\sigma_{\Delta,I\sub{int}}^2(l))$
much larger than unity indicates a good match of the
centroid $\Delta$-variance spectrum with the true velocity spectrum. 
{\changed The denominator grows} relative to the numerator with 
increasing scales when the density spectrum is steeper than the
velocity spectrum. {\changed In this case, matched by the lower panels
of Figs. \ref{fig_delta_densitytruncation}-\ref{fig_delta-decomposition},
the largest deviations of the centroid spectrum from the velocity
spectrum should occur at large scales, whereas the {\newchanged slopes
of the $\Delta$-variance spectra should match} at small
scales. This is {\newchanged indeed the behaviour} that we observe in
these figures.} For the opposite relation {\changed of spectral 
indices, where the velocity spectrum is steeper than the density spectrum,
as seen in the upper panels of the figures, the ratio is growing
towards larger scales, and in fact {\newchanged we find} the
best matches of the scaling behaviour at large scales and the}
main deviation at small scales. 

{\changed Using the $\Delta$-variance
spectra in Fig. \ref{fig_delta_densitytruncation}, we can evaluate 
the criterion by eye from the plots. When the curves for
the centroids fall well above the dotted line giving the density spectrum,
the centroid spectrum should be a reliable tracer of the velocity structure.
The same test can be performed} in the analysis
of observed data, because the {\changed $\Delta$-variance spectra of
the intensity and the centroid velocity maps} and the average velocity
dispersion are
easily measured in observed line data. However, we find that the
actual significance is also limited. In the upper panel of Fig. 
\ref{fig_delta_densitytruncation} with the shallow
density and steep velocity spectrum, {\changed we find that the velocity
spectrum is reproduced by the centroid spectrum} when the
centroid $\Delta$-variance exceeds the values from the {\newchanged column} density structure
by about a factor four, whereas for the steep density and shallow velocity
structure we get a good match even if the centroid curve falls {\changed just
above the {\newchanged column} density spectrum.} 
{\changed For other combinations of spectral indices} we find
that a ratio of
two is sufficient to guarantee a match between centroid spectrum
and velocity spectrum as long as the density spectrum is very steep
($\beta_{\rho} > 3.5$), whereas ratios as high as 100 may be required
to guarantee a match when the density spectrum has an index shallower than
2.5. {\changed When using the normalised centroids in Fig.
\ref{fig_delta-normal_densitytruncation} we cannot derive an equivalent
criterion to estimate the match between centroid scaling and velocity scaling
based on the measured map spectra.} This is a clear practical
advantage of the weighted centroids.

Thus we can basically confirm the criterion, when applied in its scale
dependent form to $\Delta$-variance spectra, but have to emphasise that
there is no single value for the ratio where the transition between
velocity-dominated and ``density-contaminated'' behaviour appears, but that
the exact shape of the density spectrum has to be taken into account.


\subsection{Comparison of $\Delta$-variance spectra and structure function}
\label{sect_sf_2}

As the structure function is related to the autocorrelation function,
the decomposition in Eq. (\ref{eq_general_decomposition}) applies
as well to the contributions to the structure function. 
\citet{Lazarian03}, however, have shown that the second term representing the
combination of density and velocity fluctuations can be further split
into two separate
contributions in terms of the structure function. As one of them
represents the pure density fluctuations, they suggested that
the structure function can undergo a transition from a velocity-dominated
spectrum to a density-dominated spectrum.

To test this behaviour we have repeated the experiments shown in
Figs. \ref{fig_delta_densitytruncation} to \ref{fig_delta-v0}
for structure functions.
In general we expect to see clear deviations from power-laws as
the projection of structure functions results always in broken power laws.
Structure functions
of 2-D projections can be represented by two asymptotic power laws:
one at small lags ($|\vec{l}|\ll z_{tot}$) having a spectral index $\beta-2$
for both shallow and steep spectra, and another one at large lags 
($|\vec{l}|\gg z_{tot}$) with a spectral index $\beta-3$ for steep spectra
and $0$ (constant) for shallow spectra \citep{Esquivel}.
Taking the general limitation of a restricted dynamic range of scales,
both in the fBm simulations and in most observed maps,  the 2-D structure
functions will always fall in the transition between the two asymptotes so that
their slope cannot be reconciled directly, preventing a direct recovery of 
the underlying 3-D statistics.
Hence, no simple inversion of the projection problem is possible.
Compared to the $\Delta$-variance spectra, the spectra of structure functions
are thus always somewhat more curved with steeper slopes at small lags
and shallower slopes at large lags, but in spite of the different analytic
decomposition of the structure function of centroid velocities
demonstrated by \citet{Esquivel}, the general behaviour is always 
very similar to the $\Delta$-variance spectra.

The measured changes 
with respect to variations of the density and velocity zero level
is also almost identical to the behaviour shown in Figs. \ref{fig_delta_densitytruncation}
to \ref{fig_delta-v0}. For the {\changed centroids obtained from
the density structure with the 3$\sigma_{\rho}$ shift-and-truncate level
we find a very good match of the structure functions of the centroids
and the projected velocity structure.} If the density dispersion,
however, is in the order of the average density the
spectra flatten with main deviations at large scales
when the density spectrum is steep and at small scales
when it is shallow. 
When comparing $\Delta$-variance spectra and structure functions in 
detail, we find that the centroid structure functions
resemble the true velocity structure always slightly better than
the $\Delta$-variance spectra. This might be partially due
to the somewhat lower sensitivity of the structure function to
changes in the power spectrum at particular scales as found
by \citet{OML}, but might indicate also a slight advantage of
the structure function compared to the $\Delta$-variance spectra
when applied to centroid maps.

There is again no transition from a velocity-matching
behaviour to a density-matching behaviour, but rather a density-contaminated
structure with a spectrum which is shallower than the true velocity spectrum.
We find as well a confirmation of the 
scale dependent criterion of \citet{Esquivel} for a match between centroid
spectrum and true velocity spectrum.
Here, the critical ratio $X^2D_{v\sub{c}}(\vec{l})/(\langle v^2 \rangle
D_{I\sub{int}}(\vec{l}))$ for a match between centroid structure function
and velocity structure function for a particular combination of spectral indices
is always somewhat smaller compared to the $\Delta$-variance spectra. For
steep density spectra, a ratio of one seems to be always sufficient, whereas
for shallow density spectra a ratio of 20 may be required to guarantee a
velocity-dominated centroid behaviour.

\section{The derivation of the velocity structure from density-contaminated centroids}

Whenever the average density of the medium is too small so that
the centroid spectrum does no longer reflect the underlying
velocity spectrum, we can deduce the true velocity spectrum from
measured centroids only {\changed when we find a way to compute the
second term in Eq. (\ref{eq_general_decomposition}) and when the last
two terms produced by the accidental correlations are negligible.
Based on the results of the decomposition shown in Fig.
\ref{fig_delta-decomposition}, we propose an iteration scheme which
computes the second term, i.e. the convolution of the two fluctuation
spectra, from the first term, i.e. the velocity projection weighted
with $\rho_0^2$, obtained in a previous iteration step, neglecting the
small contribution from the other two terms.}

As a first step to compute the fluctuation term a three-dimensional 
{\changed fluctuation structure has to be constructed which matches the
scaling behaviour of the measured {\newchanged column} density structure}. This can be done 
in the following way: From the measured $\Delta$-variance spectrum of
the density projection, i.e. the spectrum of the intensity map, we
can compute the {\changed three-dimensional $\Delta$-variance}
spectrum by de-projecting it according to the results from Sect.
\ref{sect_projection}. {\changed By translating this spectrum 
by $k=z\sub{tot}/l$ into a spherically symmetric power spectrum $P(k)$, we create} a
new fBm-like structure using this power spectrum and random phases.
This new structure should match the scaling behaviour of the 
input 2-D $\Delta$-variance spectrum. 

\begin{figure}
\centering
\epsfig{file=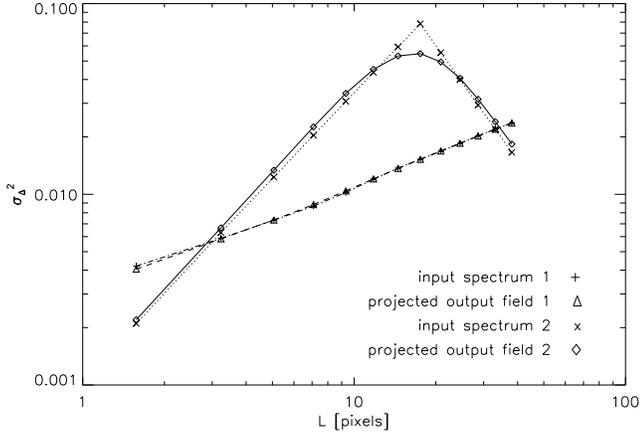,angle=90,width=\columnwidth}
\caption{Two examples for the construction of fluctuation fields
from a projected $\Delta$-variance spectrum. The figure compares the
input spectra with the spectra obtained from the $\Delta$-variance
analysis of the projection of the computed fluctuation fields.
}
\label{fig_delta2fbm}
\end{figure}

This is illustrated in Fig. \ref{fig_delta2fbm}. We demonstrate
the quality of this construction for two examples. In the first case
we use the $\Delta$-variance spectrum of the projection of a known fBm
with a spectral index $\beta=2.6$, in the second case we start
from an artificial spectrum given by a power law corresponding
to $\beta=3.7$ up to 17 pixels and an $l^{-2}$ decay above. In both
cases we create the corresponding 3-D fluctuation field, compute its
2-D projection and the $\Delta$-variance spectrum of the projection.
Comparing this derived spectrum with the input spectrum in Fig. \ref{fig_delta2fbm}
gives an impression of the {\changed quality of the reconstruction.} 

For the
power-law input spectrum we find an almost perfect match with small
deviations due to the artificial gridding of the fluctuation field,
numerical uncertainties, and statistical fluctuations.
In contrast,
the spectrum composed of two power-laws is less accurately
reproduced. The $\Delta$-variance spectrum of the fluctuation field
shows a broader peak and approaches the original spectrum only 
at lags relatively far apart from the peak. This broadening is due to the
convolution of the power spectrum with the filter function in
Eq. (\ref{eq_deltafourier}) which was ignored in the simple
translation of the $\Delta$-variance spectrum back into a power spectrum
described above. In principle we could try to include a corresponding
deconvolution to make the approach fully self-consistent, but
the reasonable agreement between the two curves even in
this extreme case shows that this additional refinement is not needed.
The example was chosen to be extreme in the sense, that we have
a sharp turn from a steeply increasing spectrum into the {\changed steep
decay of the $\Delta$-variance representing completely uncorrelated structures}.
In all cases with wider peaks, the agreement between
the original spectrum and the derived spectrum {\changed is better},
although the general tendency remains that the peak in the derived
fluctuation spectrum is always slightly too broad. The actual quality
of the construction of the fluctuation field
from the $\Delta$-variance spectrum will thus fall between the
two extremes {\changed shown in Fig. \ref{fig_delta2fbm}}.  

{\changed The fluctuation field constructed in this way has a zero average,
so that we can use it directly as the auxiliary field to compute
the centroids for $\rho-\rho_0$ in Fig. \ref{fig_delta-decomposition},
i.e. the second term in Eq. (\ref{eq_general_decomposition}). 
Unfortunately, the unknown field of velocity fluctuation enters as well
into this term so that} an iteration scheme is required: we start from the measured
centroid spectrum, assuming that it is purely determined by the
projection of the velocity field, divide by $\rho_0^2$ and construct
a fluctuation field for the velocities in the same way as 
described above for the density fluctuation field. From the 
convolution of the two fluctuation fields we estimate the 
{\changed $\Delta$-variance spectrum of the correction term. Subtracting this
spectrum} from the measured centroid spectrum then provides
the next estimate for the pure projection of the velocity
field. This can be used again to determine the {\changed
pure fluctuation term in the centroids} and so on. 
{\changed The iteration is stopped when the velocity spectrum
obtained in subsequent steps remains constant within 1\,\%\footnote{The
exact value of the convergence criterion is not important, it
only changes the number of required iterations. We found that
the results obtained for smaller error limits cannot be distinguished
by eye from the 1\,\% limit results.}}


\begin{figure}
\centering
\epsfig{file=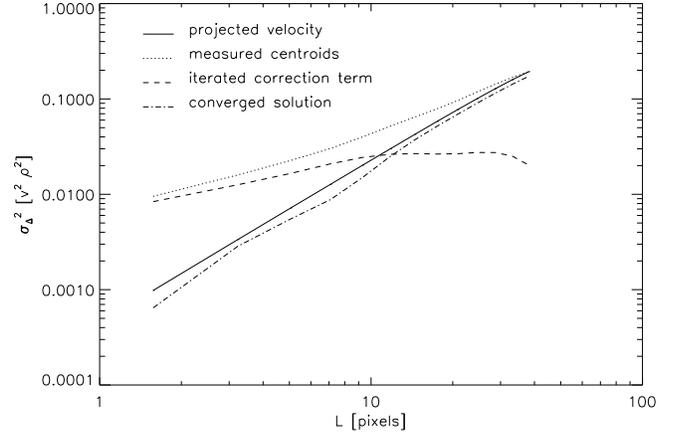,angle=90,width=\columnwidth}\\
\epsfig{file=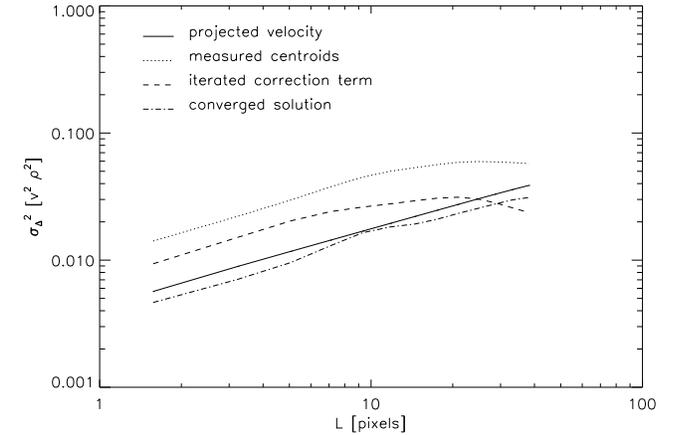,angle=90,width=\columnwidth}
\caption{$\Delta$-variance spectra of the weighted centroid velocities
for the density and velocity structures from Fig. \ref{fig_delta-decomposition}
(dotted lines).  The solid line represents the projected velocity structure with
a $\rho_0$ weighting. The dashed and the dash-dot lines represent the
correction term and the derived projected velocity contribution
at the end of the iteration. In the ideal case this converged solution
should agree with the spectrum from the original velocity structure.
}
\label{fig_iteration}
\end{figure}

An example for the result of this iteration is displayed in Fig. \ref{fig_iteration}
for the centroid spectra obtained from the combination of shallow
density and steep velocity fields and vice versa as shown in Fig. 
\ref{fig_delta-decomposition}.
The general recovery of the projected velocity structure
is quite satisfactory. The absolute magnitude of the fluctuations is,
however, somewhat too small, and for the combination of shallow density
with a steep velocity spectrum the derived overall velocity spectrum is
also slightly steeper than the original spectrum. {\changed These
remaining deviations should stem from the accidental correlations
between density and velocity field, expressed in the higher
terms of Eq. (\ref{eq_general_decomposition}).
Altogether, the}
iteration scheme has proven to be a reliable method to recover the
original velocity spectrum from a measured centroid spectrum, when
the projected density structure and the average density are known.
In all fBm combinations tested here, the overall slope of the derived velocity
spectrum agrees with that of the original spectrum within 0.1. This
is sufficient to distinguish between different turbulence models
\citep{Elmegreen}.

A major drawback of the method is the need for an accurate estimate
of the average density in the considered interstellar cloud. {\changed
This is not easy to obtain from the projected density in 
observational data because the line-of-sight extent of a cloud is
often not known.} This can be overcome in clouds with a known
geometry or by excitation studies of molecular tracers sensitive to
particular densities. However, very accurate estimates will be always
difficult.

\begin{figure}
\centering
\epsfig{file=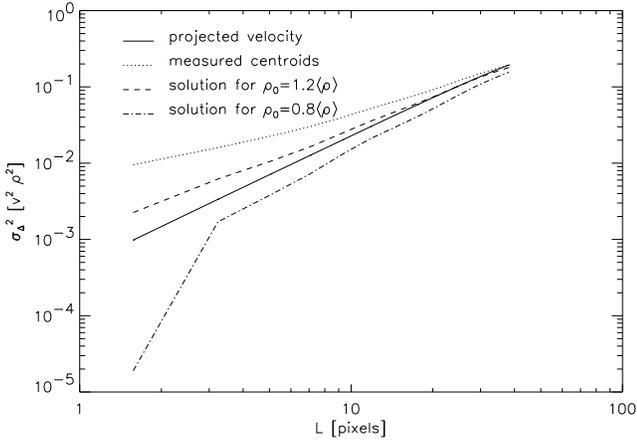,angle=90,width=\columnwidth}
\caption{Derivation of the projected velocity spectrum from
a measured centroid spectrum when applying a 20\,\% variation to the 
density $\rho_0$ used in the iteration scheme.
}
\label{fig_densjitter}
\end{figure}

Thus we have studied the influence of an error in the determination of the
average density on the reconstruction of the velocity structure. In Fig.
\ref{fig_densjitter} we have repeated the experiment shown in the upper
plot of Fig. \ref{fig_iteration} when increasing and decreasing the used
average density relative to the actual value by 20\,\%. The result shows
the same tendencies discussed in Sect. \ref{sect_zerolevel}. When the
average density is overestimated, the centroids are thought to better
resemble the scaling of the underlying velocity structure. The velocity
fluctuations are underestimated because they are obtained by dividing 
{\changed the spectrum by a $\rho_0^2$ value which is too large. The}
computed correction term is
too small and the derived velocity spectrum falls above the actual 
spectrum and is too shallow. If the average density is underestimated,
we correct the centroid spectrum with an overestimated fluctuation
spectrum, so that the derived velocity spectrum is too small and too steep.
For steep velocity spectra a change of the average density by 20\,\% corresponds
to a change of the average exponent of the spectrum by 0.25. For shallow
spectra, the influence is somewhat smaller.

Thus we can conclude that {\changed it is possible to the retrieve
the actual velocity spectrum from measured centroids even if the centroid
spectrum is density-contaminated, but the accuracy of this retrieval depends
critically on an knowledge of the average density in the cloud. Independent
measures of the $\sigma_{\rho}/\rho_0$ ratio are required. Methods 
to accurately derive the density from multi-line observations
have been successfully developed and applied by \citet[e.g.][]{Schreyer, 
Richter, Sonnentrucker}. They are based on the combination of information
from different species tracing a wide range of critical densities, but
the accurate determination of $\rho_0$ still remains a challenging task.}
\label{sect_iterate}

\section{Conclusions}

We have shown that the $\Delta$-variance analysis is an appropriate tool
to characterise the scaling properties of both velocity centroid maps
and the underlying three-dimensional velocity field. By directly
reflecting the power spectrum of fluctuations and preserving a power-law
behaviour through the projection the $\Delta$-variance is well suited to 
quantify the properties of interstellar velocity fields.
{\changed The fact that velocity centroids may not reflect the
velocity statistics was always a concern for turbulence research. The
disagreement between the aforementioned statistics was discussed already
by \citet{OML} and \citet{Brunt}). We successfully tested the criterion
for the validity of centroids as measures of velocity statistics
suggested by \citet{Lazarian03}.}

We find that the most accurate criterion determining whether a centroid
spectrum reflects the velocity scaling properties is a small ratio between
the density dispersion and the mean density. For values below 0.5 the
centroid spectra match the underlying velocity structure. Here,
the centroids are determined by the pure projection of the velocity
field. At higher $\sigma_{\rho}/\rho_0$ ratios the mutual convolution
of density
and velocity fluctuation contributes a main term. Based on this knowledge,
we can {\changed qualitatively} explain all the differences in the interpretation of 
centroid spectra found in the literature.

Without knowing the average density in the considered medium we can
test whether a centroid spectrum reflects true velocity structure
using the criterion by \citet{Lazarian03} that $X^2\sigma_{v\sub{c}}^2(\vec{l})
 \gg \langle v^2 \rangle \sigma_{I\sub{int}}^2(\vec{l})$ when
the centroid spectrum is velocity-dominated. Although derived for 
the structure function it holds for the $\Delta$-variance as well.
However, there is no single value by how much the left hand side
has to exceed the right hand side. 
We have confirmed the criterion by numerical experiments and found that factors
above two are sufficient in case of steep density spectra but factors
up to 100 may be required for extremely shallow density spectra.
Currently, observations and simulations of interstellar turbulence
show that both steep and shallow regimes may
occur with density spectral indices ranging from about 2.5 to 3.3 
(see Sect. \ref{sect_fbms}).

{\changed We do not see a transition} from velocity-dominated to
density-dominated spectra at lower densities, but rather a transition to
``density-contaminated'' spectra which are systematically shallower.
The flattening of the centroid spectra relative to the true velocity
structure in the general case can be easily misinterpreted as a transition
from a velocity-resembling to a density-resembling spectrum because
in interstellar turbulence the density spectra are {\changed often} shallower
than the velocity spectra. In any case, the density structure can be
obtained directly from column density maps. A density-dominated
spectrum occurs only if the velocity scale was chosen in an unfortunate way
so that the average velocity is not negligible relative to the velocity
dispersion. By adjusting the velocity frame such that the average
line is centred at zero, this term can always be eliminated.

Whenever the centroid spectrum is velocity-dominated, the $\Delta$-variance
analysis is a simple and robust tool to directly infer the velocity
scaling from the centroid map. The exponent of the $\Delta$-variance
spectrum is the exponent of the power spectrum of the
velocity fluctuations reduced by two. Although, the second order
structure function is connected to the power spectrum by a different
functional behaviour, we find a very similar behaviour when applied to
centroid velocities. All general conclusions apply there as well. However,
the structure functions of projections of power-law power spectra are
always curved, so that a direct fit of the exponent is more difficult.
Moreover, we find that, although our analytical decomposition of the
velocity centroids is only valid for weighted centroids,
the normalised centroids behave qualitatively in the same way
so that they can be used as well to derive the velocity structure
whenever the centroid spectrum is velocity-dominated.

We provide an iteration scheme to derive the actual velocity structure
from the centroid maps in all cases where the average density is known,
even if the map is density-contaminated. An accurate determination
of the power spectrum of the velocity fluctuations depends on three
conditions: the correlation between density and velocity structure
can be neglected, the dynamic range of length scales covered by the
map is sufficient to compensate for statistical fluctuations at
particular lags, and the average density $\rho_0$ can be estimated with
a high accuracy. 
%
%

A different iteration scheme can be developed using the structure
function instead of the $\Delta$-variance. In this case the decomposition
proposed by \citet{Esquivel} can be used to obtain a scheme which is
less sensitive to the knowledge of the average density, but a 
considerably more complex approach is needed to evaluate the projection
effects. This will be the topic of a subsequent paper.

\begin{acknowledgements}
VO was supported by  the Deut\-sche For\-schungs\-ge\-mein\-schaft
through grant 494A. AE acknowledges support from the NSF grant
AST-0307869 and the Center for Magnetic Self-Organization in Laboratory and Astrophysical Plasmas, and Mexico's Consejo Nacional de Ciencia y Tecnolog\'{\i}a. AL is supported by NSF grant AST0307869. {\newchanged We thank an anonomous
referee for comments helping to lay out many aspects in a clearer
and more precise way.}
We have made use of NASA's Astrophysics Data System Abstract Service.
\end{acknowledgements}

\end{document}